\documentclass[printer]{aa} 
\usepackage{graphicx}
\usepackage{txfonts}
\usepackage{natbib}
\usepackage{color}
\usepackage{hyperref}

\def\degr{^\circ}

\usepackage[normalem]{ulem}

\begin{document}

\title{Diffuse interstellar bands as dust indicators: the contribution from 3D maps}
%\subtitle{}
%\shorttitle{}

\author{R. Lallement\inst{1}
\and J.~L. Vergely\inst{2}
\and N.~L.~J. Cox\inst{3}
}

\institute{GEPI, Observatoire de Paris, PSL University, CNRS, 5 Place Jules Janssen, 92190 Meudon, France\\
       \email{rosine.lallement@obspm.fr}
\and 
ACRI-ST, 260 Route du Pin Montard, BP234, 06904 Sophia-Antipolis, France 
\and 
ACRI-ST, Centre d’Etudes et de Recherche de Grasse (CERGA), 10 Av. Nicolas Copernic, 06130 Grasse, France
}

\date{Received ; accepted }
\titlerunning{DIBs as dust indicators - the contribution from 3D maps}
% \abstract{}{}{}{}{} 
% 5 {} token are mandatory
 
\abstract
% context heading (optional)
%leave it empty if necessary 
{Tridimensional (3D) distributions of the 862~nm diffuse interstellar band (DIB) carrier have been computed based on {\it Gaia} parallaxes and DIB catalogues, in parallel with 3D maps of dust extinction density. 3D maps provide local diagnostic in addition to line-of-sight information, information on distributed structures, and allow to focus on poorly studied low extinction areas. They make cross-matching with other catalogs possible through estimates of DIB and extinction along any given path.}
% aims heading (mandatory}
{We aim to re-examine the relationships between the density of DIB carriers and the absorption and emission properties of spatially co-located dust. Along with laboratory identifications of carriers, these properties may shed light on the formation and evolution of this organic matter. They may also help modeling dust emission and absorption properties in a more detailed way.}
% methods heading (mandatory)
{We used the 3D maps of 862 nm DIBs and dust as well as available DIB equivalent width (EW) catalogs and published measurements of parameters characterizing the dust extinction law and the dust emission. We studied the relationships between the extinction-normalized 862~nm DIB EW and the extinction level, the total-to-selective extinction ratio, $R_V$, the dust FIR emission spectral index, $\beta$. We re-visited  the link between several DIBs and the UV absorption bump at 220~nm.}
 % results
{The ratio, DIB$_\mathrm{norm}^{862}$, between the 862~nm DIB carrier density and the optical extinction density is increasing in low density clouds, confirming with local values the trend seen in line-of-sight data. In both cases, the coefficients of a fitted power law fall within the range of those measured toward SDSS high-latitude targets for 20 different bands, ranking this DIB among those with a high increase, above the one of the broad 4430 ~\AA\ DIB. This is consistent with the recent measurement of a larger scale height above the Plane for the 862~nm DIB compared to the one of the 443 ~nm DIB. \\ Using map-integrated 862 nm DIB EWs and extinctions along the paths to APOGEE targets with published proxies $R'_V$ for the total to selective extinction ratio, we found that, despite a large scatter, DIB$_\mathrm{norm}^{862}$ is positively correlated with $R'_V$ for those stars with low to moderate extinctions ($A_V$= 0.2 to 2--3~mag). Independently, using stars from the 862~nm DIB catalogue located outside the disk and for the same regime of extinction, DIB$_\mathrm{norm}^{862}$ is found to be globally anti-correlated with the Planck opacity spectral index $\beta$. This is consistent with the observed anti-correlation between $\beta$ and $R'_{V}$. In the light of a recent result on the variability of the carbon/silicate ratio in dust grains as a source of this anti-correlation, it suggests that DIB$_\mathrm{norm}^{862}$ increases with the fraction of carbonaceous to silicate grains in the co-located dust, in agreement with the carbonaceous nature of DIB carriers and recent evidences for a spatial correlation between DIB$_\mathrm{norm}^{862}$ and the fluxes of carbon-rich ejecta of AGBs. At higher extinction, both trends disappear, and there is evidence for a reversal. \\ About the link between the height of the 220 nm UV absorption bump and extinction-normalized EWs of DIBs, we found that two factors explain the absence of previous, clear results: the correlation disappears when we move from ($\sigma$-type) DIBs to $\zeta$-type ones and/or from single-cloud lines of sight to paths crossing multiple clouds distant from each other. $\zeta$-type bands can be used to predict low and high values of the bump height, provided one adds a correcting factor linked to the ambient radiation field, e.g. the 5780/5797 DIB ratio. We show examples of  simple models of the bump height based on the 5780 ~\AA\ band, the 5850~\AA\ band and the 5780/5797 DIB ratio. We also found an anti-correlation between DIB$_\mathrm{norm}$ and the width of bump, similarly disappearing from $\sigma$-type to $\zeta$-type DIBs. This suggests that a fraction of the bump is generated outside the dense molecular clouds.}
{There are complex relationships between DIBs and dust, however, massive measurements of DIBs and extinction and the derived 3D maps may provide some constraints on the density, the nature, and the contribution to extinction and emission of the co-located dust. This requires large stellar spectroscopic surveys and space-based measurements of UV extinction.}
%Data on several selected strong DIBs from massive spectroscopic surveys and space-based measurements of UV extinction would help building DIB and dust maps, understanding the intricate evolution of DIB carriers and dust grains and refining models of dust absorption and emission spectra.}

%list of allowed keywords: https://www.aanda.org/author-information/information-files/170-aaa-keywords

%\keywords{Diffuse Interstellar Bands; Interstellar Medium; Interstellar dust; AGB stars}

\maketitle

% Diffuse interstellar bands (DIBs) have been observed with increasing details and for an increasing number of targets for the last ten years. In particular, the Radial Velocity Spectrometer (RVS) on board Gaia has allowed the production of extended catalogs of equivalent width (EW) measurements for the 862~nm diffuse interstellar band (DIB), along with accurate target star distance. 
% Gomez-Munoz 204 : fullerene-rich PNs: absorption in FUV rise fitted with HAC support idea that fullerenes from HAC by top down in PN envelopes

\section{Introduction}\label{intro}

Diffuse Interstellar Bands (DIBs) have fascinated (and, also, sometimes discouraged) already two generations of specialists of the interstellar medium and astrochemists. These weak, irregular absorptions are imprinted in the spectra of all astrophysical sources located behind interstellar clouds. About 600 of these features of various intensities and widths have been observed in the visible and near infrared, and discoveries of new bands are continuously made, especially for faint or very broad bands \citep{Ebenbichler24, Maiz21}, and in the NIR \citep{Hamano16}. 
%xxxx add Castellanos24 if accepted
Their width, shape, and strength strongly vary from one band to the other (for catalogs and recent reviews, see \cite{Jenniskens94, Hobbs2008, Geballe11, Cox14, Hamano16, Cox17, Fan19}. Their multiple-peak substructure favors large carbonaceous molecules or molecular ions \citep{Sarre95}, but only one of these carriers could yet be identified, the buckminster fullerene cation C$_{60}^{+}$ \citep{Foing94, Campbell15, Cordiner19}. 

For all DIBs, there is a strong link between the DIB strength and the extinction along the line of sight, as well as the total amount of gas. However, there is a significant scatter on these relations. At least two influencing factors have been identified. First, the intensity of the radiation field \citep{Vos11, Friedman11} acts on almost all carriers, and in a different way from one to the other. Second, there is a significant decrease of the DIB strength relative to the amount of dust for lines of sight crossing the most opaque regions of the clouds \citep{Krelowski88, Herbig95, Cami97, Vos11, Elyajouri17}, the so-called skin-effect. But other factors certainly enter in play, as shown by \cite{Ensor17}. 

Although they probably constitute a large reservoir of interstellar organic matter, little is known about their formation, thought to be either gradual around evolved stars together with dust and other carbonaceous complex particles such as PAHs \citep[see, e.g.,][]{Kwok23}, or in dense molecular clouds along with other identified complex molecules. In support of the former mechanism, C$_{60}$ and C$_{70}$ have been detected in emission in planetary nebulae \citep[PNe, first detections by ][]{Cami10}. However, enhancements of DIBs (C$_{60}^{+}$ or others) along lines of sight towards PNe and attributable to the PN itself have been too scarce to establish a link between DIB carriers and PNe.
% (xx Zasowski red square nebula but not confirmed for optical DIBs, also this nebula is not a classical PN).
More recently, evidence has been reported for enhanced density of 862~nm DIB \footnote{For this particular DIB observed with {\it Gaia}-RVS we use the nm notation, while we keep the traditional Angstr\"om notation for other DIBs.} carriers in regions of the Milky Way harboring AGBs with strong carbon-rich dust emission \citep{Cox24}. Because C-rich AGBs ejecta enrich the ISM in carbonaceous matter, favoring the formation of carbonaceous macromolecules, while O-rich ejecta favor the formation of silicates, such a link may appear natural. However, it does not necessarily imply the formation of the DIB carrier within the ejecta. It may simply enhance the carbon abundance in the region, which in turn impacts on all types of molecular processes.

One may expect some continuity between the macromolecules at the origin of DIBs and the smallest dust grains \citep[see][ for a review]{Jones16}. This is why links between DIBs and other phenomena potentially related to carbonaceous matter, free-flying molecules or grains have been searched for, such like the 220~nm {\it bump} feature, the extended red emission (ERE), the Unidentified Infrared Emissions (UIEs). Unfortunately, tight relationships between DIB EWs and dust extinction parameters could not be found, because there are only very few stars for which the path from the Sun to the target is strongly dominated by a unique, homogeneous cloud. The situation is even worse for links with dust emission data, because, as an additional limitation, only stars located beyond the dust layer are appropriate targets.
Links between DIB carriers and the UV bump have been searched since decades and in different ways. See \cite{Xiang11} for an exhaustive review of the many attempts. Among the positive results, \cite{Krelowski92} suggested a dependence of the normalized DIB on the {\it bump}, based on six early-type, highly reddened stars. Later on, in a full spectral analysis of 28 stars, \cite{Desert95} found a positive correlation between the DIB to reddening ratio of a few strong DIBs and the height of the 220~nm {\it bump}, as well as, in parallel, an anti-correlation with the width of the bump and with the FUV rise assigned to nano-grains. More recently, \cite{Xiang11} disfavored dependencies on the basis of too weak correlation coefficients, and \cite{Xiang17} confirmed this negative result. \cite{Xiang17} also did not find any correlation between DIBs and the far-UV rise of the extinction curve. As part of a study of potential relationships between the DIB carriers and cometary material, and based on the compilation of \cite{Xiang17} data, \cite{Bertaux17_dibs_comets} argued that high values of the well-known ratio between  DIB 5797~\AA\ and DIB 5780~\AA\, or 'radiation-shielding' ratio have some link with the steep FUV rise of the line-of-sight extinction law. They interpreted this trend as a coincidence between the increase of ultra-small grains in dense cloud cores and the disappearance of the 'radiation-resilient' 578.0~nm DIB. More recently, \cite{Ramireztannus18} and \cite{Li19} investigated the link between several DIBs and the total to selective extinction ratio $R_V = A_V$ / E($B$-$V$), based on the same high extinction lines of sight towards M17 ($A_V$ $\geq$ 4). \cite{Ramireztannus18} found a negative correlation between the DIB EW normalized to the reddening E($B$-$V$) and $R_V$ (they also included the peculiar/extreme line of sight toward Herschel\,36), while \cite{Li19} found no correlation between the DIB EW normalized to the extinction $A_V$ (instead of the reddening) and $R_V$. In summary, the link between DIB strengths and the extinction law is far from clear. As far as dust continuum emission is concerned, to our knowledge nothing has appeared in the literature linking DIBs to emission parameters. Potential relationships are more difficult to establish, DIB being distance-limited quantities, while emission parameters characterize entire lines-of-sight. 

As already mentioned, tight relationships between DIBs and dust extinction or emission parameters can not be found, and only trends may be detected. However, we believe that such trends may contain valuable information. The goal of the present work is to search for such trends using new {\it Gaia} data and 3D maps based on this data. Namely, the availability of DIB equivalent width measurements for massive amounts of stars distributed in 3D space has allowed to perform their inversion to derive local values, i.e. quantities proportional to volume densities of the carriers. This was recently done by \cite{Cox24} using the extended 862~nm DIB catalog of \cite{Schultheis23} based on RVS spectra. In parallel, and importantly for our present work, \cite{Cox24} performed a specific inversion of extinctions based on the same lines-of-sight as those used for the DIB inversion, and for the same parameters of the inversion code. This way, the two 3D DIB and extinction maps are suitable for optimal comparisons. 

The existence of 3D density maps brings several new possibilities. Firstly, it allows the comparison between local values of DIB carriers and of other interstellar constituents (in particular dust grains responsible for the extinction). Such relationships are more direct diagnostics than those based on Sun-star integrated values. They may reveal links with the environment, the radiation field, or the presence of specific types of stars. As an example, the above-mentioned evidence for a link between the 862 nm DIB to extinction ratio DIB$_\mathrm{norm}^{862}$ on the one hand, and the presence of carbon-rich Asymptotic Giant Branch stars (AGBs) on the other hand, was found based on the 3D maps. Secondly, the maps allow to focus on dust-poor areas, overlooked in studies based on line-of-sight (LOS) data. A third advantage of a 3D map is that it constitutes a tool for estimates of integrated quantities along the line-of-sight to any target, including those which were not part of the inverted catalog used to produce the map, and have been observed for various purposes. Finally, a fourth advantage is the possibility of a documented selection of regions and directions appropriate to a specific study, e.g. selections of single-cloud lines-of-sight or targets beyond the layer of dust. In this work, we made use of all these advantages. 

The 862 nm DIB is a broad ($\simeq$ 0.4 nm wide) DIB. It is not particularly strong, however, being spectrally close to ionized calcium (CaII) lines commonly used to constrain the stellar parameters, its study benefits from stellar spectroscopic surveys. It was intensively studied based on the spectra of the RAVE survey and was been found to be well correlated with the reddening \citep{Munari08,Kos13,Kos14}. This was also confirmed by data from the {\it Gaia}-ESO and GIBS surveys \citep{Puspitarini15, Zhao21}, and the 862 nm DIB was found useful as a tracer of the arms and of their dynamical properties. It was shown to be well correlated with the NIR DIB at 1.5273 $\mu$m \citep{Zhao21}. It did not show any particular characteristics that set it apart from other bands, and, as a result, it is probably representative of most other DIBs.

Section~\ref{sect1} is devoted to the relationship between the 862~nm DIB and the extinction, and to the comparison with DIB-extinction relationships found for other DIBs. In Sect.~\ref{sect2} we detail the search for a link between the total-to- extinction ratio $R_V$ and the DIB strength normalized to the extinction (DIB$_\mathrm{norm}^{862}$). In Sect.~\ref{sect3} we similarly detail  the search for a link between DIB$_\mathrm{norm}^{862}$ (for the 862~nm DIB) and the dust opacity spectral index $\beta$ measured by Planck. In Sect.~\ref{sect4} we revisit the studies between DIBs and the 220~nm {\it bump} of the extinction law. Section~\ref{Conclusion} discusses the various types of relationships, the limitations and their potential use. We also discuss some implications.

%We used available DIB catalogs and 3D maps to investigate relationships between DIB strengths, extinction, total-to-selective extinction ratio, dust emission spectral index beta, and the extinction 220~nm {\it bump}. 
%To do so, a selection of sightlines for which there is negligible interstellar dust beyond its extremity is necessary, because the spectral index refers to the totality of the dust in the chosen direction. We found a negative correlation between DIB$_\mathrm{norm}^{862}$ and beta (dust) for Av$\leq \simeq$ 2.5 mag, and a reversal of the trend for high extinction above this threshold. The trend at low-moderate Av is more marked when using recent estimates of beta by \cite{Casandjian22}. The two trends, for Rv and beta are found fully compatible with the average negative correlation between beta and Rv from \cite{Schlafly16}, which reinforces their likeliness. 
%We made use of the 3D 860.2nm DIB and extinction maps to compute DIB$_\mathrm{norm}^{862}$ for all sightlines for wich Rv values were estimated by \cite{Schlafly16}. We found a positive correlation between DIB$_\mathrm{norm}^{862}$ and Rv for the reddening regime Av $\leq$ $\simeq$ 2.5 mag. We show that at high extinction the trend is reversed, in agreement with the results of \cite{Ramireztannus18} for Av $\geq$ $\simeq$ 4 mag.

\begin{figure*}[th!]
   \centering
    \includegraphics[width=0.9\textwidth]{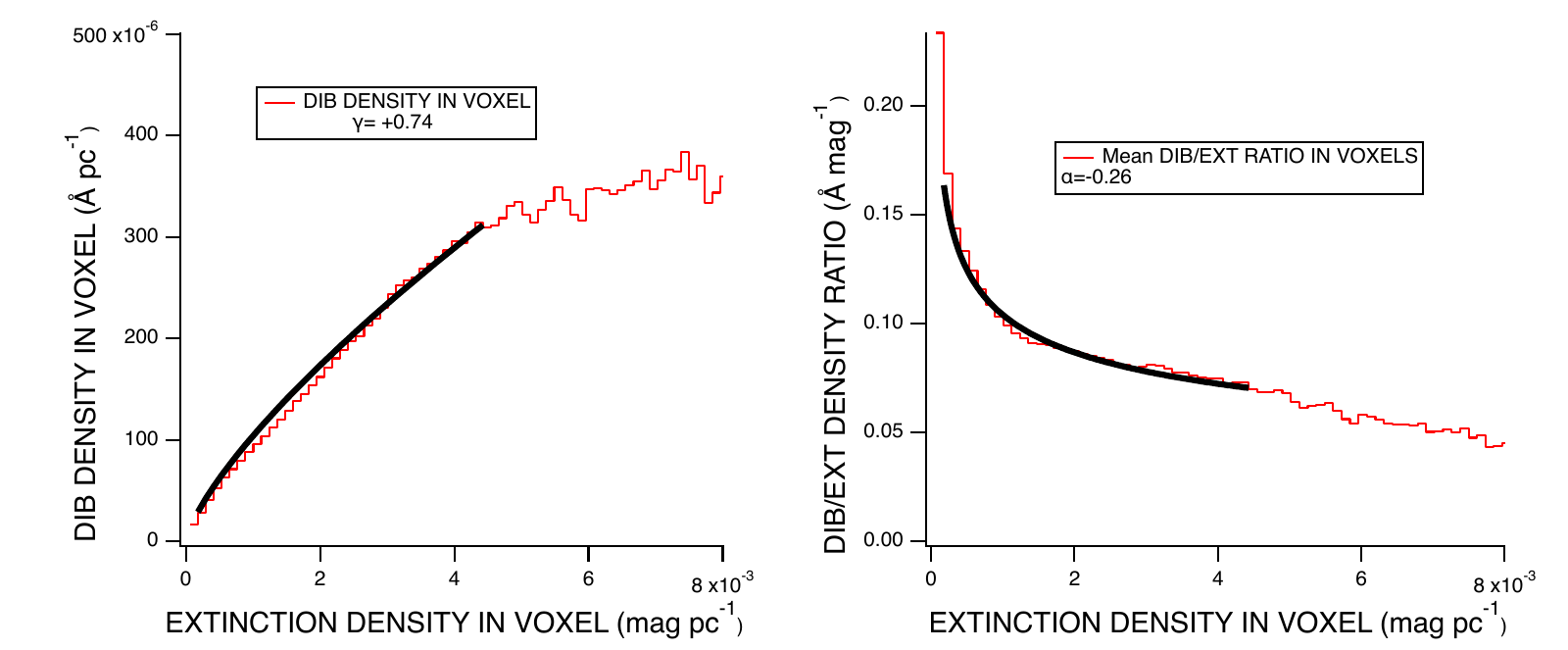}
    \caption{DIB carrier and extinction volume densities. All voxels of the \cite{Cox24} 3D maps have been used. Left: DIB density averaged in voxels with equal extinction density. An adjusted power-law limited to the low extinction regime is superimposed (see text). Right: Ratio between the DIB EW density and the extinction density for the same voxels, and superimposed power-law.}
    \label{DEPENDENCE_EXT}
\end{figure*}

\begin{figure}[th!]
   \centering
    \includegraphics[width=0.45\textwidth,height=6cm]{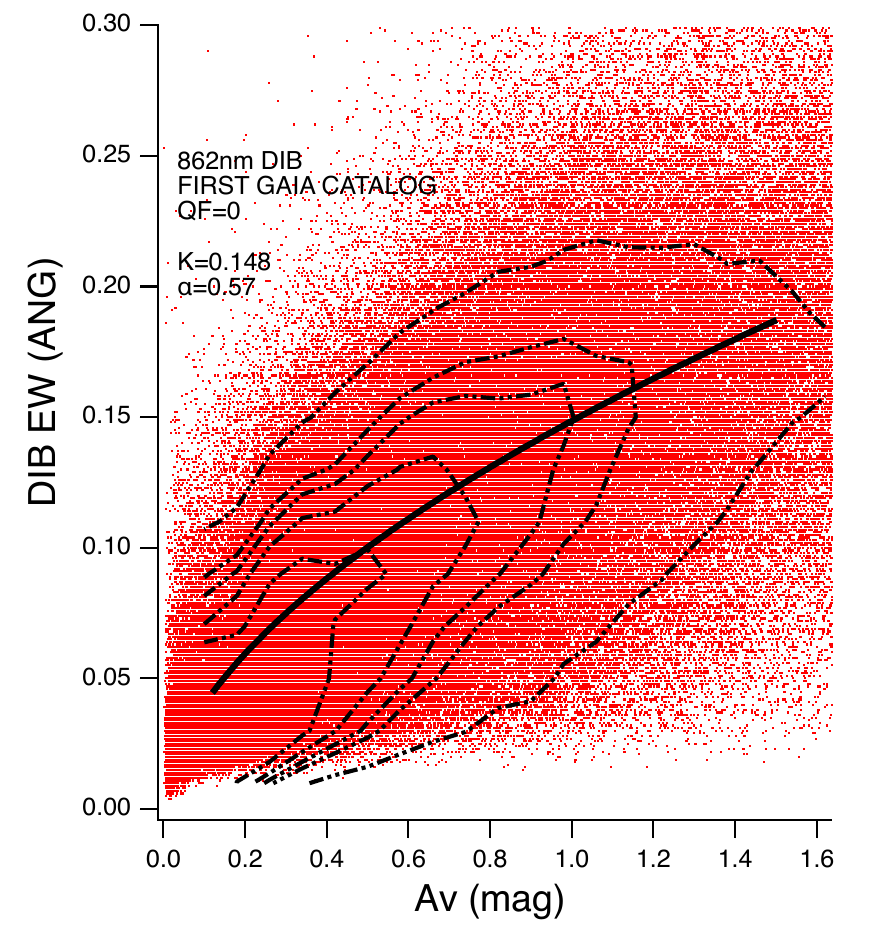}
  \includegraphics[width=0.45\textwidth, height=6cm]{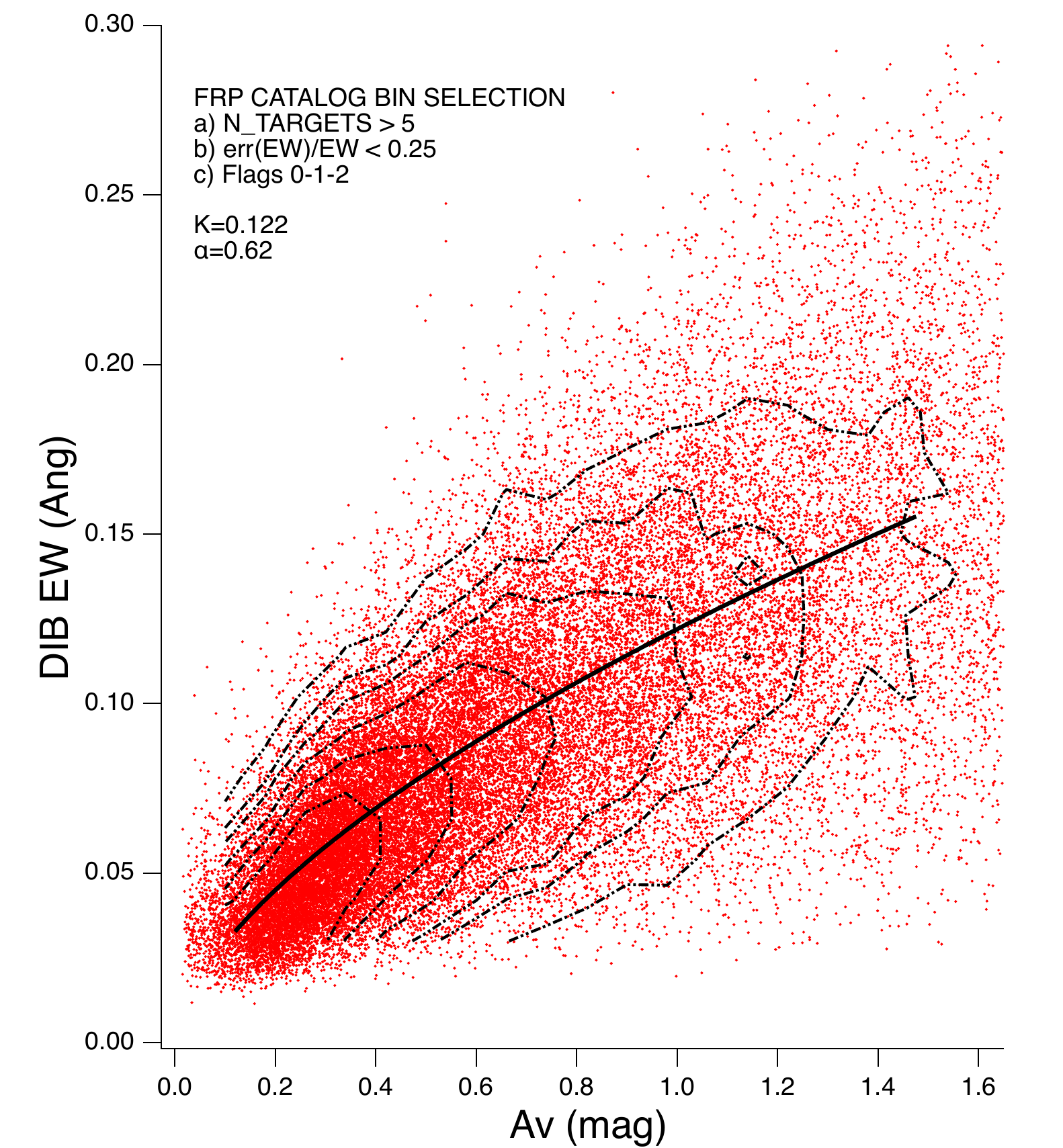}
 \caption{Left: DIB EW from the {\it Gaia} DR3 catalog as a function of the extinction Av integrated from the Sun to the target star. A power-law is fitted to the data from Av=0 to Av=1.5 mag. Iso-contours of target number densities have been drawn to illustrate the various contributions. Right: same as left, for the second catalog of averaged EWs in spatial bins.}
    \label{DR3_FRP_DIB_EXT}
\end{figure}

\section{Extinction-normalized DIB and extinction} \label{sect1}

%xxx insert Schlamrmann21 : C60+ 9632-9577/EBV also decrease with ebv > same type of behavior than 8620 corresponds to low gamma de Lan 
 %5780 5797 6203 suggest decrease at very low ebv but does not agree with Lan (also 1) nothing for 6203 based 5797 and 5780 based on two points...
%NOTE: a tros/4 etoiles a low ebv et low EW/EBV en deduit descente a low ebv pas confirme par Lan et surtout 2 spec binary 1 emission line et une nebuleuse voir igor
%par contre plat jusqu'a EBV=1.1
%confirme Lan pas d'effet comme DIB GAIA ou C60 ou autres

%xxx. here local bubble stuff?).
Significant attention has been given to DIB variability in star-forming regions and, in particular, to DIB carrier depletion in dense, UV protected parts of dense clouds (the {\it skin effect}). It is due, in a large part, to the preference for highly extinguished target stars. Conversely, much less attention has been given to the low extinction part of the DIB dependence on dust opacity. The 3D maps of the 862~nm DIB \citep{Cox24} do not possess a high enough spatial resolution to address the skin effect for individual clouds. However, for the first time, it is possible to use the maps to investigate the relationship between the local value of the DIB carrier density $\rho$DIB, measured as the spatial gradient of the DIB EW, and the local value of the extinction density $\rho A_V$ in wide, low dust areas, or similarly the relationship between the extinction-normalized DIB density $\rho$DIB$_\mathrm{norm}^{862}$=$\rho$DIB/$\rho A_V$ and the extinction density $\rho$A$_{V}$. 

To do so, we selected all $10 \times 10 \times 10$~pc$^{3}$ voxels of both DIB and extinction 3D maps containing enough target stars to perform the local inversion of DIB EWs \citep[see details in][]{Cox24}. We distributed the DIB densities $\rho$DIB in these voxels in bins of similar local extinction density $\rho A_V$, and computed their average in each bin. Note that densities $\rho A_V$ are smaller than actual densities in moderate or high density clouds, because the spatial resolution of the used 3D maps is poor, including the one of the dust map because it has been especially computed based on the same series of LOS than those of the DIB catalogue, and there is a dilution of clumpy structures into the map voxels. 
We did the same sampling for the ratios $\rho$DIB / $\rho A_V$. Fig.~\ref{DEPENDENCE_EXT} displays the two distributions of bin-averaged DIB density $\rho$DIB on the one hand, and bin-averaged ratio $\rho$DIB / $\rho A_V$ (i.e. the averaged extinction-normalized DIB density $\rho$DIB$_\mathrm{norm}^{862}$) as a function of the local extinction density $\rho A_V$. Superimposed are fitted power laws: 

\begin{eqnarray}
\begin{aligned}
\rho \mathrm{DIB} & = K\, (\rho A_V)^{\gamma} \\
\rho \mathrm{DIB}_\mathrm{norm} & = \frac{\rho \mathrm{DIB}}{\rho A_V} = K\, (\rho A_V)^{\gamma-1} = K\, (\rho A_V)^{\alpha}
\end{aligned}
\end{eqnarray}

We restricted the adjustment to the extinction density interval $0.00018 < \rho A_V < 0.0044$ mag\,pc$^{-1}$. The lower limit ensures the exclusion of uncertain ratios corresponding to low statistics, while the upper limit excludes voxels containing dense clouds. The power law coefficients $\gamma$ are found to be $\gamma = +0.74$ for the DIB vs extinction dependence, and, accordingly, $\alpha = \gamma-1 = -0.26$ for the normalized DIB dependence. 

In addition to local values, we also used directly measured line-of-sight integrated values EW from two catalogs, the first catalog of individual lines of sight \citep{Schultheis23} and the second catalog of averaged EWs among stars belonging to the same spatial bin \citep{Schultheis23_frp}. For both two datasets, we used the EXPLORE 3D extinction density maps \citep{Vergely22} to compute the corresponding integrated extinctions, i.e., from the Sun to each target star in the first case, and from the Sun to each center of the spatial bin in the second case. Fig. \ref{DR3_FRP_DIB_EXT} displays the EW measurements for individual targets of the first catalog (left), and for spatial bins from the second catalog (right), both as a function of the corresponding integrated extinction. 
An adjusted power law is superimposed for the two datasets. We restricted the fit to the low-moderate extinction interval $0.12 \leq A_V \leq 1.5$~mag. The $\gamma$ coefficients were found to be 0.57 and 0.62 respectively for the two catalogs. The two coefficients are similar, and are slightly lower than the one found from 3D maps (0.74). Taken as a whole, these results provide an order of magnitude of the uncertainty on the coefficients for such a simple relationship. One can see that they depend on the DIB extraction method and they also differ depending on the use of direct or inverted measurements. However, the existence of similar trends for both integrated values and local values confirms that the increase, for lower values of the dust extinction density, of the extinction-normalized DIB strength is well established. Especially, the result based on local values, taken from 3D distributions, confirms that it is not due to an existing bias in the measurements of weak extinctions, or low DIB EWs. 
%xx explain the telescopic mode xx. xxx needs more discussion xxx why this difference? I would have expected the opposite...smoothing when doing maps? JL? 
%This result is to compare with the dependence found for the same DIB by \cite{Damineli16} for line-of-sight integrated quantities, namely DIB(int)=$K' A_{Ks}^{\gamma'}$ and $\gamma'$ =0.83. Both agree about a significant increase of the DIB carrier density in dust-poor areas (although Damaneli based on high extinctions 

%XXXX maybe not relevant already skin effect averaged since Munari finds quasi linearity check..) . The weaker increase found for integrated quantities is not surprising, since by construction there is a some averaging along the lines-of-sight. 

%xxcheck des papiers gaia 
 %dib frp intercept a eux fig21 ebv=-0.03 seulement mais pas de points en dessous de E(BP-RP)=0.2 donc EBV=0.1 difficile de voir evolution
 % dib gaia premier papier fig 7 on voit clairement que dib ne descend pas a zero aussi vite que E(BP-RP) ca pourrait clairement avoir un intercept assez grand si fit lineaire comme Lan region ebv jusqu'a 0.6 
 % pour avoir une idee un coeff gamma de 0.75 et un fit sur ebv 0.1 0.6 ca fait intercept a -0.13 
% mais coeff de 0.75 pour valuer locale ca doit faire coeff plus proche de un pour valeurs integrees

The interesting aspect about this exercise is the comparison with the results obtained by \cite{Lan2015} for a series of different DIBs. The authors used Sloane Digital,Sky Survey (SDSS) spectra of halo stars, galaxies and quasars at high Galactic latitudes and extracted 20 different DIBs from stacked spectra in a large fraction of the high-latitude sky. They adjusted power-law functions to the DIB EWs as a function of the reddening, for the same extinction interval as the one we used above in the case of the {\it Gaia} catalog (namely 0.12 $\leq$ $A_V$ $\leq$ 1.5~mag). Their table 3 lists power law coefficients $\gamma$ for the DIB dependence on E($B$-$V$). The whole set of $\gamma$ coefficients is in the range from 0.51 to 1.13 and six of the DIBs have a low $\gamma$ between 0.51 and 0.74. This demonstrates that the 862~nm DIB behaves like several other DIBs, but is part of those bands with the highest relative increase with respect to extinction in dust-poor regions (i.e., the lowest $\gamma$). By comparison, for the strong 5780 ~\AA\ DIB \cite{Lan2015} derived a $\gamma$ coefficient close to 1, which explains the linear correlation between its EW and the extinction in the regimes of low to moderate extinction. A comparison can also be made with the 957.7 and 963.2~nm C$_{60}^{+}$ DIBs. \cite{Schlarmann21} inferred from published data a strong decrease of more than a factor of three of the DIB/reddening ratio when the E(B-V) reddening is increasing from 0.1~mag to 0.6~mag (quantities derived from their figure 3c). This would correspond roughly to a power law coefficient $\gamma$ $\simeq$ 0.4, i.e. lower than the lowest from the \cite{Lan2015} series (0.51). The stability of the C$_{60}$ structure may explain why it can persist in low density, generally UV-permeated clouds. 

Since the density of dust decreases with the distance to the Plane, the $\gamma$ coefficient is expected to have some link with the DIB carrier scale height, namely, the lower $\gamma$, the larger the scale height. A recent study by \cite{Zhao23} derives a significant difference of scale heights for two DIBs, namely the 443~nm DIB and the 862 nm DIB, the 862 nm DIB scale height being larger than the one of the 443~nm DIB. The \cite{Lan2015} $\gamma$ coefficient for the 443~nm DIB is 0.89, above our three estimates of $\gamma$ for the 862 nm DIB, all three below 0.74. This is consistent with the heights measured by \cite{Zhao23}. Further measurements of DIB scale heights should help confirming or not their direct link with the $\gamma$ coefficient. We note that the hierarchy of DIBs with respect to these \cite{Lan2015} $\gamma$ coefficients does not seem to correspond to the hierarchy of $\zeta$ to $\sigma$ DIBs, related to the UV radiation field \citep[see, e.g., the work of][]{Ensor17}. This means that the phenomenon responsible for the increase of carrier density in low dust clouds is not directly or simply related to the $\zeta$ vs $\sigma$ classification. It deserves further study, being, instead, a likely indication of the carrier formation process and site. 
% the first is linked to the top-down building of carriers , the second on accretion , coagulation xxx

%xxx extend actually a compilation of all 5780 5797 or 6284 would be useful find Gaia target stars close to dib stars and compare 
%In the case of the integrated measurements, a comparison can be made with
%This relative increase in dust-poor regions is in some way linked with the ambient UV radiation field pervading in those regions, being partly correlated with DIB ratios used as indexes of the field such like the 5780/5797 or 6284/5797 ratios. 

%-normalement 4 relations maps dr3 et frp plus visees indiv dr3 et frp bins
%-fitter les quatre comparer avec lan
%-comparer avec apogee?
%-montrer 5780 pente 1 
%-voir aussi dib gaia ges

%-montrer diff bulle locale?

\section{Extinction-normalized DIB and total to selective extinction ratio R$_{V}$}\label{sect2}

%xxxx INSERER PEUT ETRE IMPORTANT
%d'abord ramirez-Tannus pentes tres forcees par etoiles a disques!!!!
%les DIBs a forte pente avec R sont dib Gaia, et 4430 (et les deux C60+)
%pour les premieres ce sont celles qui ont un faible gamma chez lan mais on a vu avec Schlmarmann que c'est aussi le cas pour 9577-9632
%aussi pentes nulles avec R pour celles qui ont un gamma grand chez Lan
%Lien entre resilience/formation dans milieu diffus et subsistance dans les disques????
%Trouver l'article recent qui ne trouve pas de dib dans les disques...
%xx ne pas oublier zhang lamost Rv map pas de catalogue donc pas de comparaison

\cite{Schlafly16, Schlafly17} published a large catalogue of measurements of a proxy $R'_V$ for the total to selective extinction ratio $R_V = A_V$ / E($B$-$V$), estimated for the lines-of-sight to SDSS/APOGEE target stars. To do so, they combined accurate stellar atmospheric parameters derived from APOGEE spectra with available photometric data. They found wide-area variations in $R'_V$ on large angular scales ($\geq$30$\degr$). The lack of link of these variations with the interstellar (IS) matter column density led the authors to conclude that they are not governed by grain size changes in dense clouds and are due to other effects acting on large scales. They also noticed that $R'_V$ is globally higher outside the solar circle (at Galactocentric distances larger than the one of the Sun). Interestingly, a global increase of the extinction-normalized 862~nm DIB with the Galactocentric radius is also one of the results of the 3D mapping presented in \cite{Cox24}, a radial increase potentially attributed to the increased prevalence of ejecta from C-rich AGBs over ejecta from O-rich stars. If the two phenomena were linked, it would imply the existence of a positive correlation between the extinction-normalized 862~nm DIB EW DIB$_\mathrm{norm}^{862}$ and $R'_V$, which motivated the following study. 
 
 \begin{figure*}
  \centering
    \includegraphics[width=0.9\textwidth]{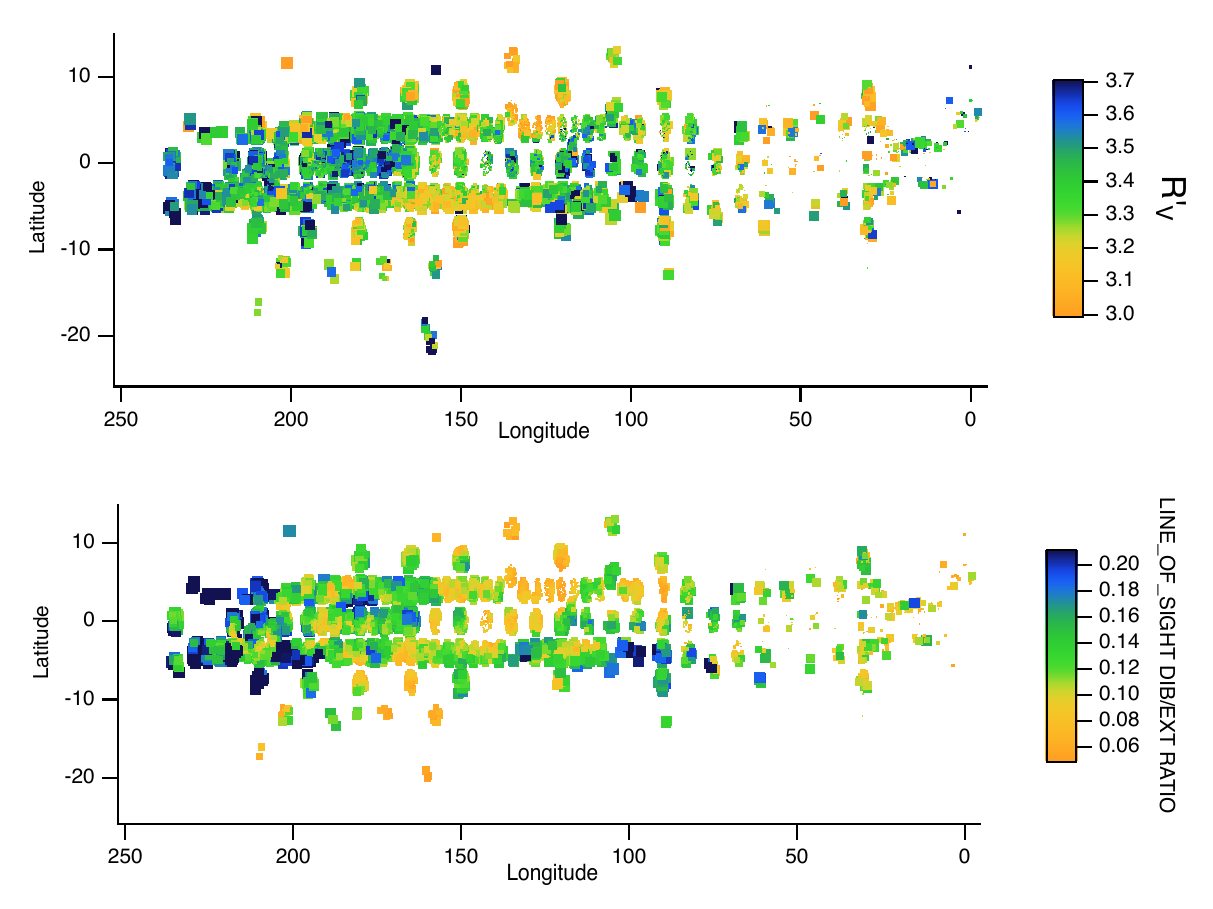}
    \caption{$R_V$ values for APOGEE targets from \cite{Schlafly16} (top) and extinction-normalized DIB for the same lines of sight (bottom) in Galactic coordinates. Sizes of the markers were reduced for line-of-sight extinctions beyond 4~mag. The limited resolution of the 3D maps is responsible for the slightly smoother image for the normalized DIB. There is a some similarity between the two independent measurements, with notable exceptions, like in the Aquila region ($l = 0\degr$, $b = +5\degr, +15\degr$) , or Taurus ($l = 160\degr$, $b = -15\degr$ or $b = -22\degr$) characterized by a strong opacity.}
    \label{Sky_dibnorm_beta}
\end{figure*}

\begin{figure*}
   \centering
    \includegraphics[width=0.9\textwidth, height=11cm]{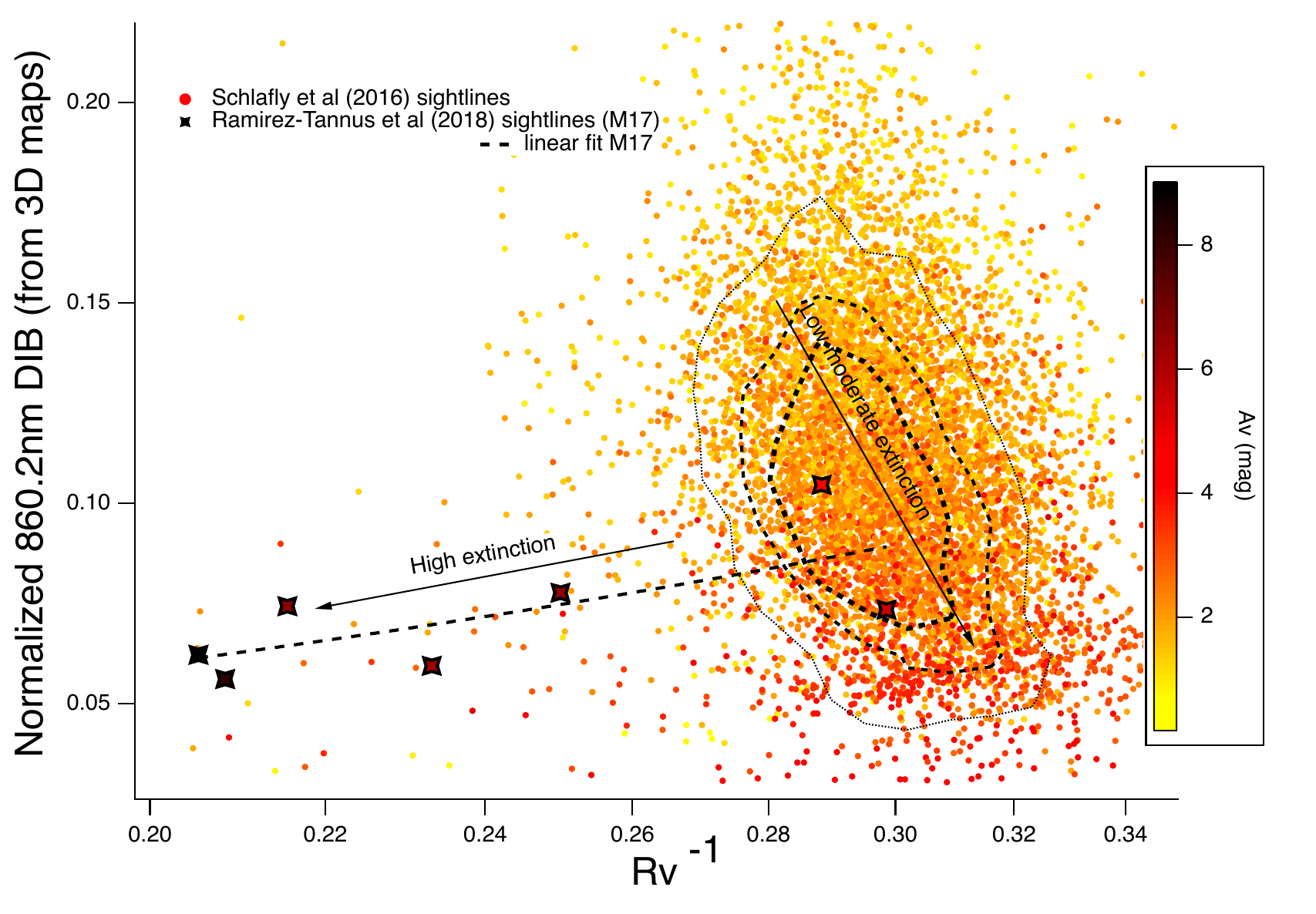}
    \caption{Relationship between the extinction-normalized 862 nm DIB and the \cite{Schlafly16} total to selective extinction R'$_{V}$ for APOGEE lines of sight (dots). For low to moderate extinctions, and, despite large scatter, DIB$_\mathrm{norm}^{862}$ decreases with R'$_{V}^{-1}$. To illustrate the trend, iso-contours of data point numbers per 0.007 $\times$ 0.04 rectangular bins in R'$_{V}^{-1}$ $\times$ DIB$_\mathrm{norm}^{862}$ are drawn for 40, 80, 120 points. The trend is reversed at very high extinctions, as shown by superimposed results of \cite{Ramireztannus18} for M17 (thick crosses). For these latter data points we neglected the difference between R'$_{V}$ and R$_{V}$. All markers are color-coded according to the LOS extinction.}
    \label{FigS16R18}
\end{figure*}

We used the catalog of \cite{Schlafly16} and for all lines-of-sight to the APOGEE targets, we integrated along these paths within the 3D distribution of DIB density from \cite{Cox24} and within the 3D distributions of extinction density from \cite{Vergely22}, to get estimates of the 862~nm DIB cumulative EW on the one hand, and of the extinction Av of the target on the other hand. Their ratio DIB$_\mathrm{norm}^{862}$ was then computed. Since the spatial extent of the \cite{Vergely22} 3D maps is resolution-dependent, we used the series of extinction maps in a hierarchical mode. We started the integration in the highest resolution map, and, if the map boundary is reached before the target star location, we continue to integrate in the second best resolution map, and so on. Targets outside the 3D DIB density map, of smaller extent  than the widest extinction map (+/- 4~kpc instead of +/- 5~kpc), were eliminated. As said in section 1, this is one of the potential uses of 3D maps, namely the computation of DIB and extinction estimates for any target star being not part of the inverted catalog. The APOGEE catalog is well appropriate for our study for the following reasons: it is massive, it is homogeneously obtained, most of the targets are distant but the extinction is limited, i.e. very dense cloud cores are generally avoided. The last aspect is important, since the resolution of the 3D maps is too poor to reconstruct very small structures. Instead, they are appropriate for the study of large-scale characteristics. 

\begin{figure}
   \centering
    \includegraphics[width=0.49\textwidth]{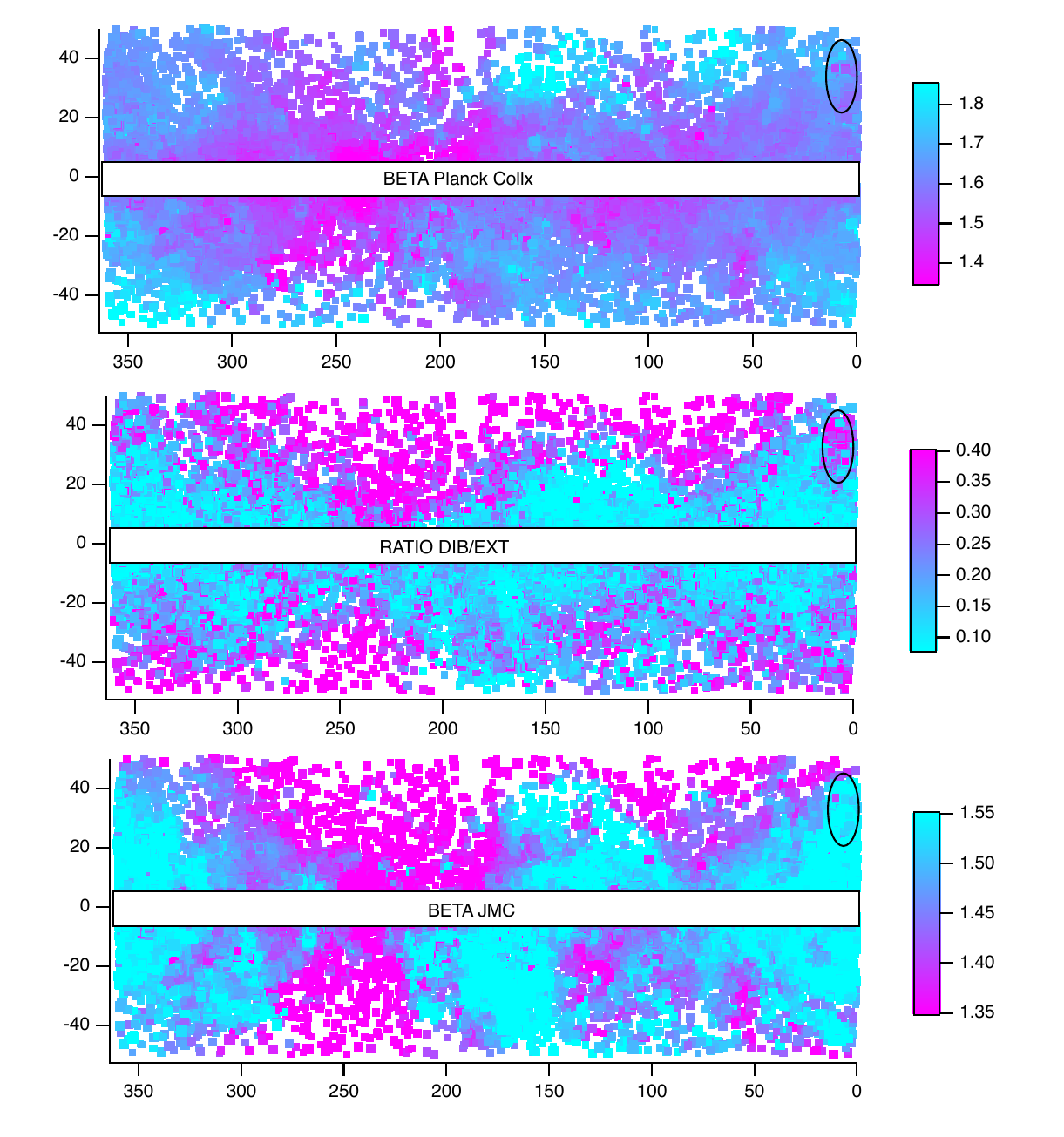}
    \caption{Images of 862~nm DIB$_\mathrm{norm}^{862}$ (middle) and the two estimates of the dust opacity spectral index $\beta$ (top: \cite{Planck2016} and bottom: \cite{Casandjian22}). There are strong similarities, again with notable exception in high extinction regions, mainly the Aquila rift area (black ellipse).}
    \label{Figskytwobetas}
\end{figure}

We made a first visual comparison of distributions over the sky of $R'_V$ measurements and computed extinction-normalized DIB EWS DIB$_\mathrm{norm}^{862}$. Fig.~\ref{Sky_dibnorm_beta} shows the locations of the APOGEE lines of sight in Galactic coordinates, color-coded according to $R'_V$ from \cite{Schlafly16} (top) and to the normalized 862 nm DIB (bottom). There are some large-scale similarities, as, e.g., in the third quadrant where both quantities are particularly high. There are also discrepancies, especially in the highly opaque directions above the Galactic centre, or in the Taurus area, where $R'_V$ is high but the DIB to extinction ratio is notably low. As a result, these sky distributions suggest a link between the two quantities, but certainly different regimes must exist according to the extinction. We note that the $R'_V$ values from \cite{Schlafly16} vary between 3.0 and 3.7, while DIB$_\mathrm{norm}^{862}$ changes by a much larger factor on the order of three. 

A more direct comparison is presented in Fig.~\ref{FigS16R18}, where the extinction-normalized DIB for each star located within the 3D DIB map is displayed as a function of the inverse of the total-to-selective extinction ratio proxy R'$_V^{-1}$. Since all extinctions that have been inverted to produce the 3D extinction maps were estimated under the assumption of a unique, fixed R$_{V}$=3.1 parameter, to avoid a bias and take into account the actual variability of R$_{V}$, we multiplied each individual extinction by a factor $R'_{V}$/3.1 before computing the DIB to extinction ratio. Despite the very large scatter, there is a clear decrease of DIB$_\mathrm{norm}^{862}$ vs R'$_{V}^{-1}$ (i.e., an increase with R'$_{V}$). This decrease occurs in the low to moderate extinction regime A$_V \lessapprox$ 3~mag (extinctions are color-coded), which is the range for most APOGEE targets used by \cite{Schlafly16}. Apart from the largest group of data points, one can distinguish several higher extinction data points at the lower left of the graph (R'$_{V}^{-1}$ between 0.2 and 0.26) which strongly depart from the general trend. This suggests that the trend disappears for some directions characterized by a high extinction. Indeed, the decrease vs R'$_{V}^{-1}$ observed at low to moderate extinction, is opposite to what was found by \cite{Ramireztannus18} for R$_{V}$ based on targets of the M17 open cluster. However, the regime of extinction for the M17 stars considered by the authors is clearly above the one of APOGEE targets (A$_{V}$ varies between 5 and 9 mag for the M17 targets). We have superimposed their results in Fig.~\ref{FigS16R18}, using the same color code representing the extinction as the one used for APOGEE data. The M17 data points depart from the main distribution of APOGEE data points to reach lower $R_V$$^{-1}$ and simultaneous low DIB$_\mathrm{norm}^{862}$ values. From the combination of the two results, it appears that the DIB-$R'_V$ positive relationship is suppressed when the extinction is reaching levels of the order of $\simeq$3 mag, with the appearance of the negative trend noted by \cite{Ramireztannus18}. 
%We could not perform the same type of study for the $R_V$ map of \cite{Zhang23} based on LAMOST data since they are not available, however, we note that the authors found a significant decrease of $R_V$ for sighlines such like those to the nearby Taurus dense molecular clouds 
%XXXX if one removes the highest point of Ram-Tan (CS disk) then slope almost zero
High $R_V$ values are generally associated with large grains formed by accretion or coagulation in UV-shielded molecular clouds, with high extinctions and DIB depletion. Indeed, this is the trend observed by \cite{Ramireztannus18} above $A_V$ = 3~mag. On the other hand, the opposite trend appearing at lower extinction is a different phenomenon. 

\begin{figure*}
   \centering
    \includegraphics[width=0.95\textwidth, height=8cm]{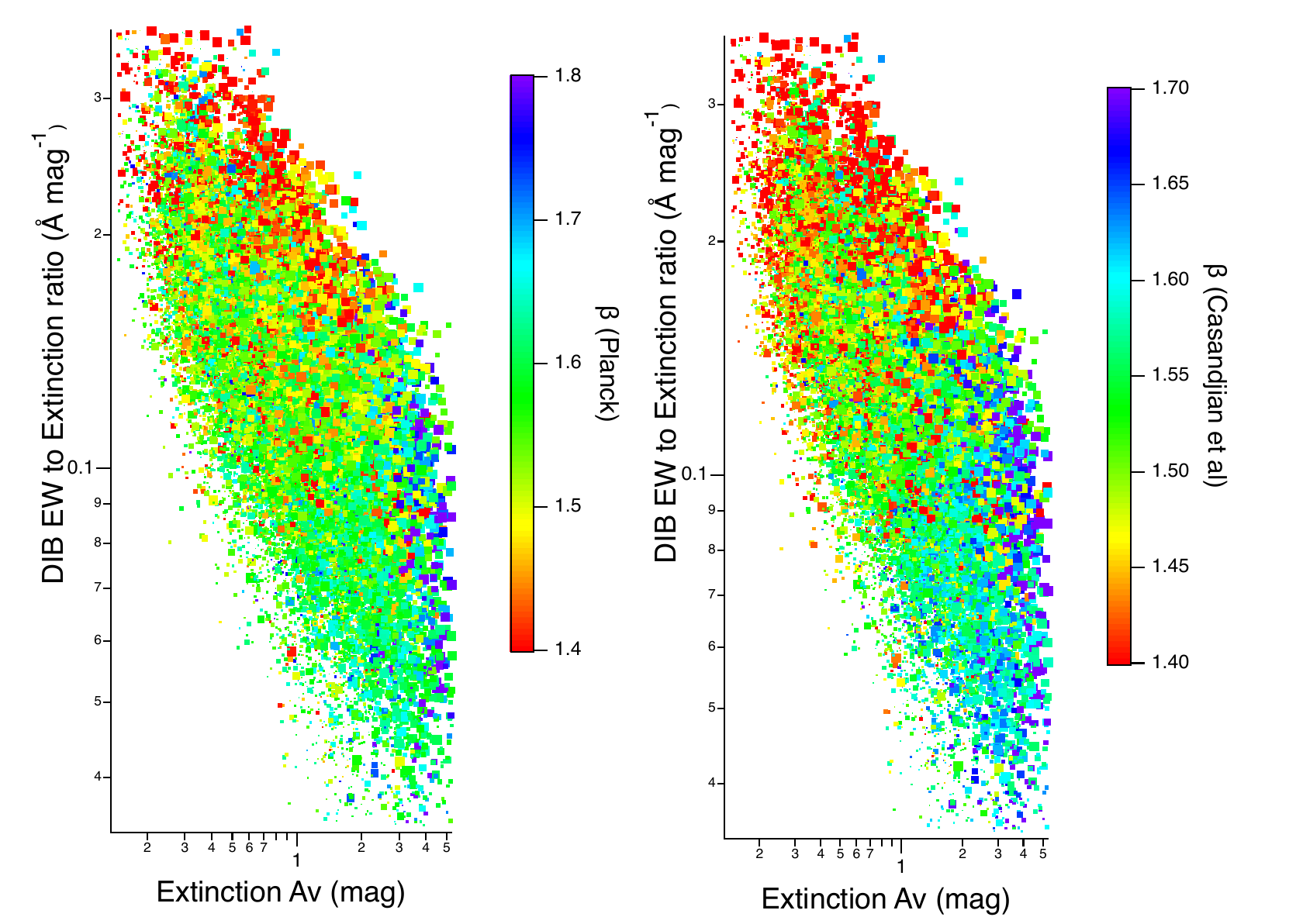}
    \caption{Illustration of the dependence on the extinction regime of the relationship between the normalized DIB and $\beta$. Plotted are the normalized ratios as a function of the LOS extinction, color-coded for the $\beta$ value for each LOS. Marker sizes increase with target location distance to the Plane. The two estimates for $\beta$ are shown (left and right). Below Av=$\simeq$ 3mag (left part of the graphs), decreasing DIB$_\mathrm{norm}^{862}$ (vertical descending lines) corresponds to an increase of $\beta$ (color from red to green). Above $\simeq$ 3~mag (right part of the graphs), there is an opposite trend: decreasing DIB$_\mathrm{norm}^{862}$ (again vertical descending lines) corresponds to a decrease of $\beta$ (color from violet to green or cyan).}
    \label{figdibnorm_vs_ext_colorbeta}
\end{figure*}

\begin{figure}
   \centering
   \includegraphics[width=0.49\textwidth]{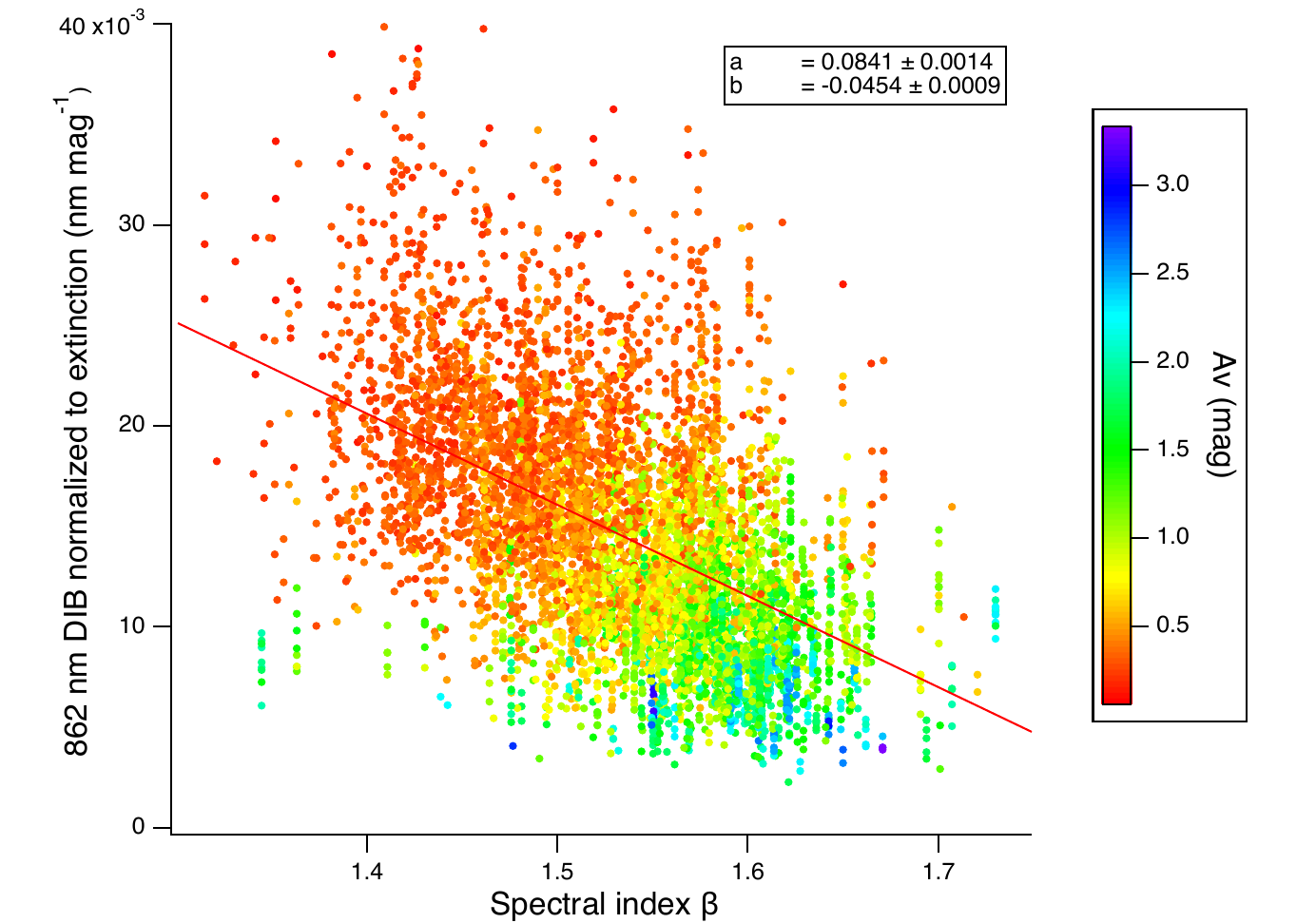}
  \caption{Extinction-normalized 862~nm DIB EWs as a function of the spectral index $\beta$ from \cite{Casandjian22}, for DIB flag zero data from the second catalog of \cite{Schultheis23_frp}, and Galactic latitudes between 10$\degr$ and 20$\degr$. Superimposed is a linear fit for this catalog selection.}
    \label{figdibnorm_vs_beta_latitudes10_20}
\end{figure}

\begin{figure}
   \centering
    \includegraphics[width=0.49\textwidth,height=8cm]{ 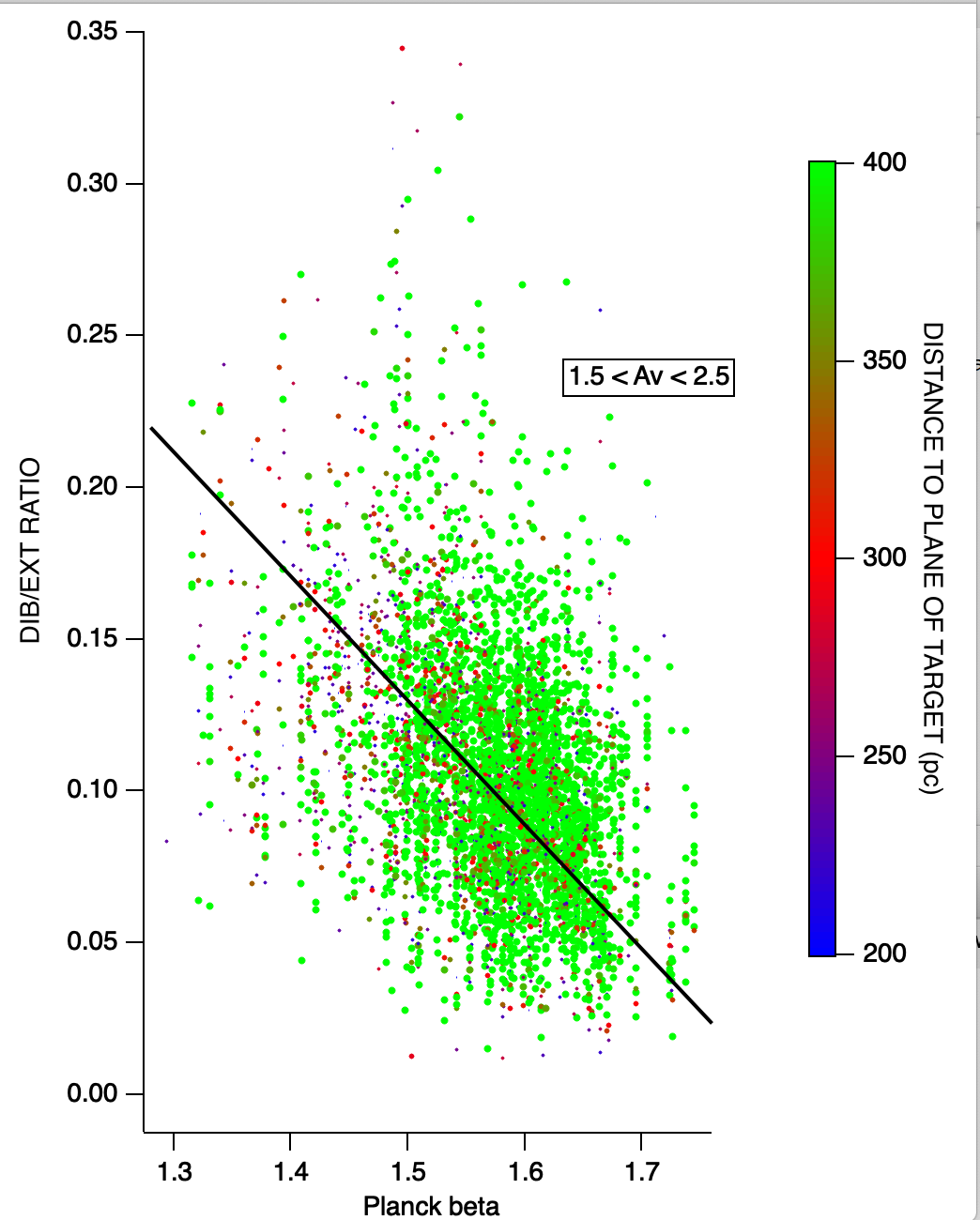}
    \caption{Extinction-normalized 862 nm DIB EW for selected data from the second catalog \citep{Schultheis23_frp} as a function of the sky-interpolated Planck spectral index $\beta$ (see text for selection). The color scale refers to the distance to the Plane of the spatial bin, and excludes distances below 200 pc. Extinction values are restricted to the 1.5 to 2.5~mag interval. An orthogonal distance regression linear fit is superimposed (see text).}
    %CHECK  (see dib frp all.pxp) selection myselecbis (((errel inf 0.5) and (ew8620 sup 0)) and ((flags8620 inf 4) and (ntargets sup 0) )) 
    %et avant au depart 235428 points Normalized DIB as a function of beta for moderate extinctions. }
    \label{figdibnorm_vs_beta_moderate_av}
\end{figure}

\begin{figure}
   \centering
    \includegraphics[width=0.49\textwidth,height=7cm]{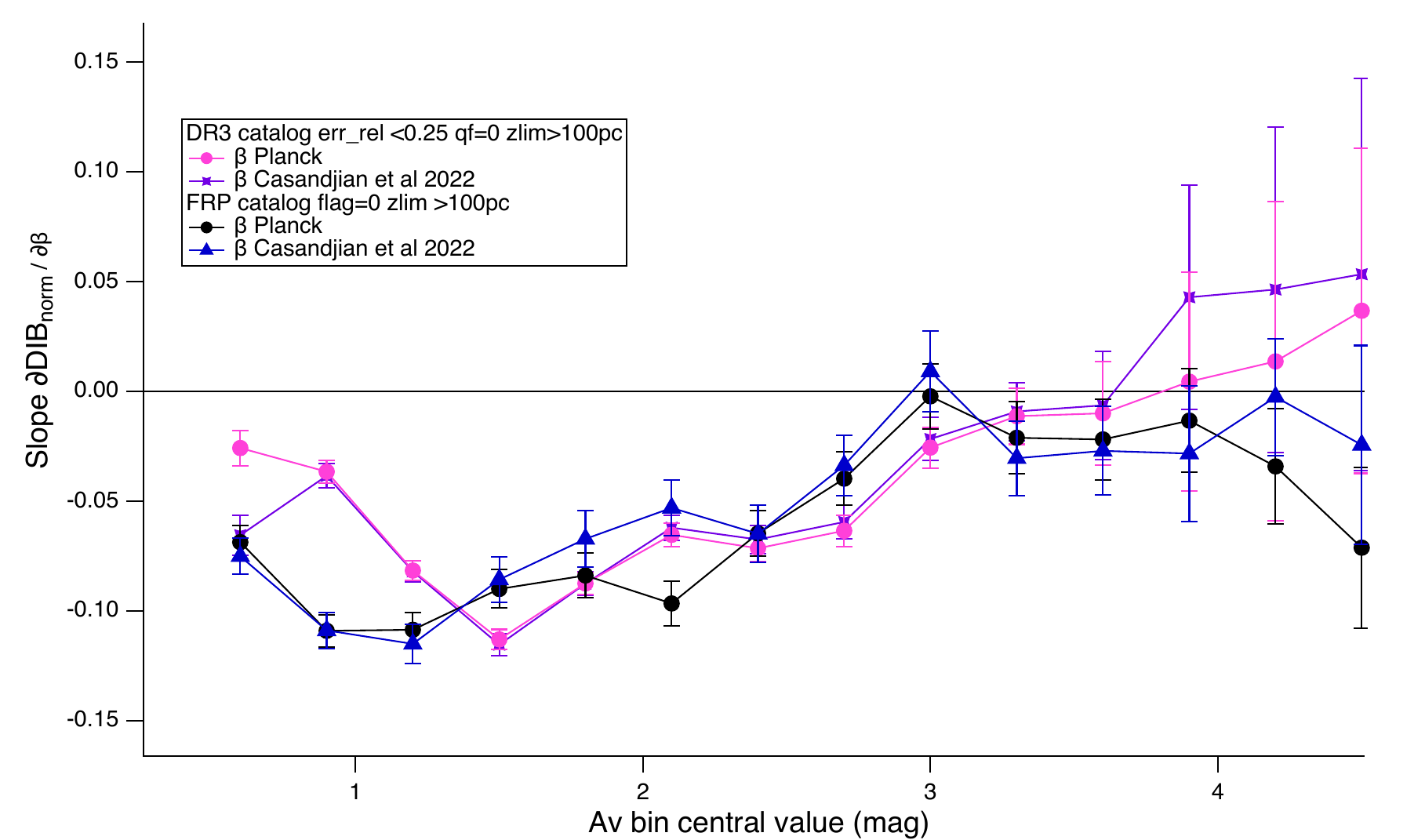}
    \caption{Ranges of gradients for extinction-normalized DIB EWs as a function of $\beta$ for data from the two catalogs and in various regimes of LOS extinctions. Errors are 1$\sigma$ uncertainties on the slopes. There are differences at low extinctions due to the very different types of data (individual lines of sight vs spatial bins) and at high extinctions due to the small number of data points. Between 1 and 3 magnitudes of extinction, all determinations are compatible and show a negative slope.}
    %CHECK (see dib frp all.pxp) selection myselecbis (((errel inf 0.5) and (ew8620 sup 0)) and ((flags8620 inf 4) and (ntargets sup 0) )) 
    %et avant au depart 235428 points Normalized DIB as a function of beta for moderate extinctions. }
    \label{fig_slopes_dibnorm_vs_beta}
\end{figure}

\section{Extinction-normalized DIB and dust emission spectral index from Planck}\label{sect3}

In the frame of their study, \cite{Schlafly16} discovered a tight inverse relationship between $R'_V$ and the dust opacity spectral index $\beta$, the parameter used in the modified black-body representation of the Galactic dust FIR emission. Properties or mechanisms at the origin of this trend have been searched for and, more recently, \cite{Zelko20} found that increasing the carbon to silicate ratio in dust grains may explain simultaneous an increase of $R_V$ and a decrease of $\beta$. This effect comes in addition to the known increase of $R_V$ in opaque clouds, attributed to the increased size of the grains. Such a result on the influence of the dust composition has important consequences in the context of extinction laws and also cosmological observations, through dust effects on the CMB spectral distortion \citep{Zelko21}. If the above positive correlation between DIB$_\mathrm{norm}^{862}$ and $R'_V$ for low-to-moderate extinction APOGEE lines-of-sight is due to the existence of a physical link between the DIB carrier density and the grain composition, it should, in turn, be accompanied by a negative correlation between DIB$_\mathrm{norm}^{862}$ and the FIR dust emission spectral index $\beta$. This expectation motivated the following study. 

As already mentioned, contrary to R$_V$, $\beta$ is not a distance-limited quantity and target stars or locations in the 3D maps must be carefully selected to ensure that most of the IS matter in their direction is located in front of them, and there is negligible IS matter beyond. Here we used the first catalog of individual 862~nm DIBs of \cite{Schultheis23} due to the more extended distribution of directions and selected DIB data for which the quality flag is 0. For each direction we estimated the spectral index $\beta$ by interpolation in the high resolution Planck map \citep{Planck2016}. We also used, in parallel, the more recent map of \cite{Casandjian22}. For each target, we computed the extinction A$_V$ by integrating in the highest resolution dust extinction density maps from \cite{Vergely22}, available from the EXPLORE facility. Similarly to the comparison with R$_V$, we first show, for a preliminary visual comparison, the distributions over the sky of the DIB to extinction ratio DIB$_\mathrm{norm}^{862}$ and the two interpolated $\beta$ values (Fig.~\ref{Figskytwobetas}). We masked the Galactic latitudes lower than 6$\degr$, since in those directions $\beta$ is too strongly influenced by dust beyond the limits of the extinction maps. We tuned the color coding of the $\beta$ maps in order to obtain about the same contrast in $\beta$ and DIB$_\mathrm{norm}^{862}$. There are obvious similarities of the patterns at large angular scale. We note that the optimal similarity is somewhat higher in the case of the \cite{Casandjian22} estimates of beta. In both cases, there are also obvious discrepancies, as e.g. at $\simeq$5-10$\degr$ Galactic longitude and $\simeq$ 20$\degr$ - 30$\degr$ Galactic latitude in the Aquila Rift region.

Fig.~\ref{figdibnorm_vs_ext_colorbeta} illustrates the complex dependence on $\beta$ of the extinction-normalized DIB, and, especially, the variability of this dependence according to the extinction regime. Here we used data from the second catalog of \cite{Schultheis23_frp} for its reduced uncertainties on the EWs and selected flag=0 data. The extinction was computed by integration in the 3D maps from the Sun to the center of the spatial bins. The figure displays DIB$_\mathrm{norm}^{862}$ as a function of the line-of-sight opacity. For each line-of-sight the dust emission spectral index $\beta$ interpolated from the Planck maps for the corresponding direction of each DIB bin center is color-coded. For Av values below about 3 mag, a decrease of the normalized DIB (going down along vertical lines in the figure) corresponds to an increase of $\beta$ (color variation from from red to green). Above Av around 3-4 mag, this trend is no longer visible, and there seems to be a weak reversed trend, namely $\beta$ is decreasing (color from violet to blue), however, the number of lines of sight becomes low, something expected from the rare cases of high extinction and high Galactic latitude. This bi-modal behavior, namely an anti-correlation between DIB$_\mathrm{norm}^{862}$ and $\beta$ for low to moderate extinctions, and a null-positive correlation with $\beta$ in the high opacity regime, is reminiscent of what we found for the relationship between the normalized DIB and $R_V$. Since the $R_V$ and $\beta$ data sources are completely independent, and since, as noted above, it has been shown that $R_V$ and beta are anti-correlated \citep{Schlafly16}, this reinforces the two detected trends. 

Given the complexity of the dependence on $\beta$, we performed comparisons in selected, limited regimes of Galactic latitudes and opacity. As an example of the various comparisons, Fig.~\ref{figdibnorm_vs_beta_latitudes10_20} shows the dependence on $\beta$ of the DIB to extinction ratio for Galactic latitudes between 10$\degr$ and 20$\degr$ and again for selected data (flag zero in the catalogue). The negative trend is clearly seen. Fig.~\ref{figdibnorm_vs_beta_moderate_av} shows another example of the dependence on $\beta$ of the DIB to extinction ratio for a restricted range of opacity, namely extinctions $A_V$ comprised between 1.5 and 2.5~mag. There is a significant negative slope for the DIB$_\mathrm{norm}^{862}$-$\beta$ relationship in this regime of extinction. An orthogonal distance regression linear fit is superimposed. It assumes errors of 0.01~\AA\,mag$^{-1}$ on DIB$_\mathrm{norm}^{862}$ and of 0.02 on $\beta$.

Finally, Fig.~\ref{fig_slopes_dibnorm_vs_beta} summarizes series of results of the type of those shown in Fig.~\ref{figdibnorm_vs_beta_moderate_av}, here for 0.3~mag extinction intervals distributed between $A_V$ = 0.5 and $A_V$ = 4.5~mag, for the DIB two catalogs, and for the two $\beta$ maps. At high extinction, the slope becomes very uncertain due to scarce data. At low extinction, the difference between individual lines of sight and spatial bins has a strong effect. Between 1 and 3 magnitudes of extinction, the different determinations are all compatible and show a negative slope of the DIB$_\mathrm{norm}^{862}$ vs $\beta$ relationship. In summary, our results suggest that, despite a wide dispersion around the general trend, the hitherto little studied regime of low to moderate extinction is characterized by opposite dependencies between DIB$_\mathrm{norm}^{862}$ and $R_V$ and between DIB$_\mathrm{norm}^{862}$ and $\beta$. Due to the very strong variability of the DIB to extinction  ratio associated with the extinction density (section 2), these trends are hard to detect unless massive datasets are available. 

\section{DIBs and 220 nm UV absorption {\it bump} }\label{sect4}

\begin{figure}
   \centering
   \includegraphics[height=5.2cm]{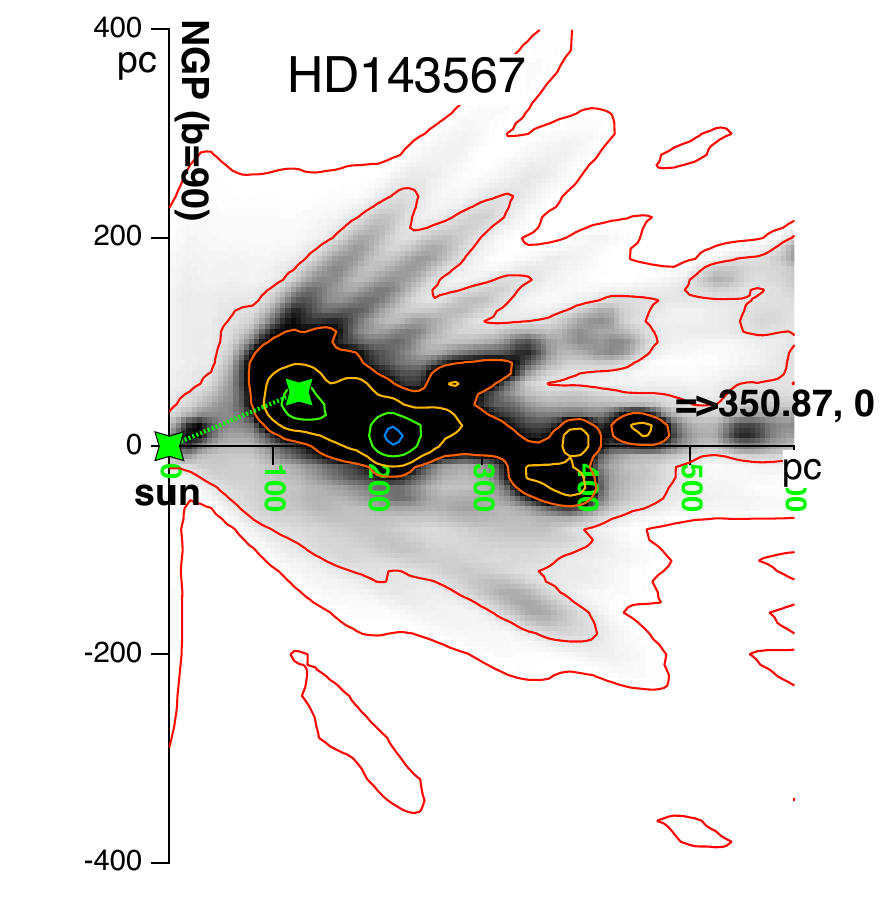}
     \includegraphics[height=5.2cm]{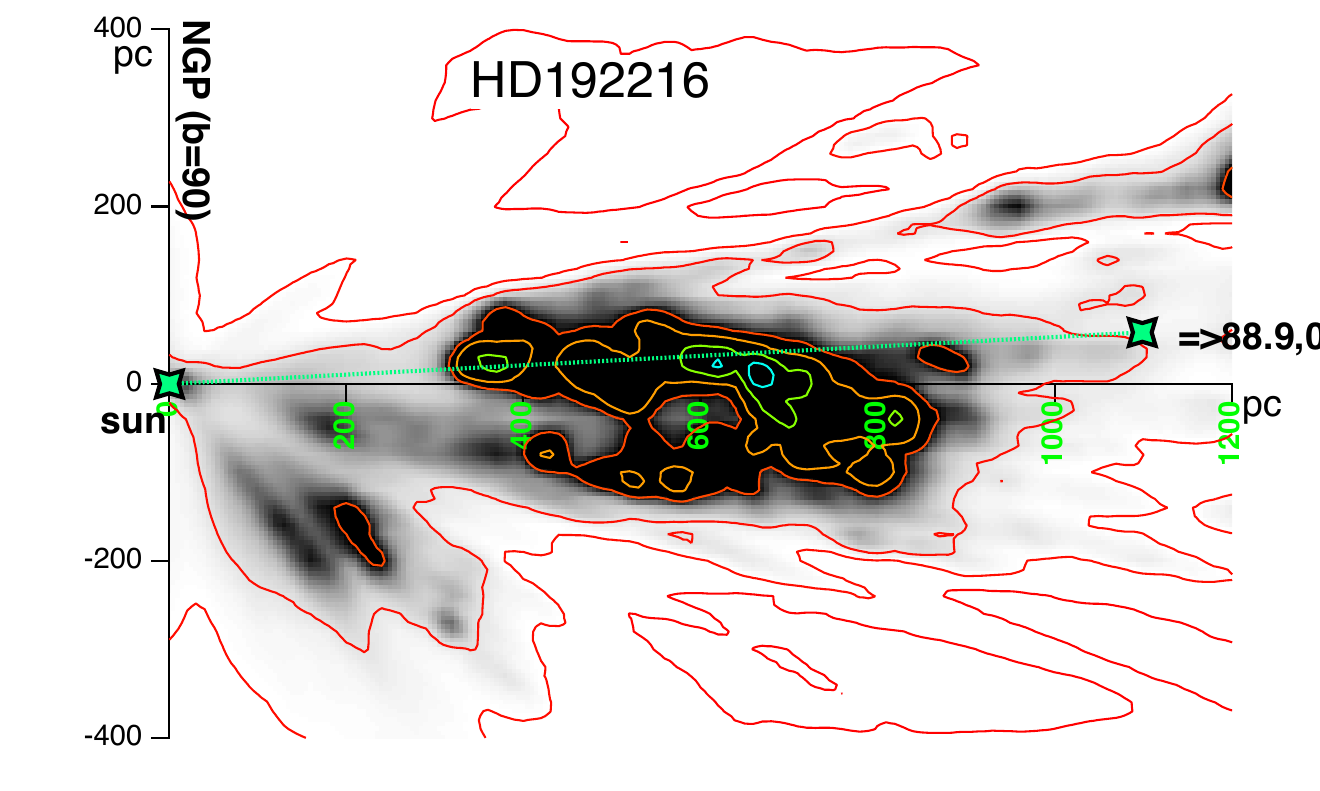}
   \caption{Illustration of the attribution of a flag to a given star from the \cite{Xiang17} list. Shown are images of the dust distribution in vertical planes containing the Sun and the star. Dark areas correspond to high dust opacity. The line of sight to the target (green marker) is shown in green. HD\,143567 (left) is attributed a flag F=1, since most of the extinction is generated in a single, very dense and very nearby cloud. The more distant HD\,192216 (left) is attributed a flag F=0, since at least two dense clouds separated by a large distance contribute to the extinction.}
    \label{visu_flags}
\end{figure}

As mentioned in Sect.~\ref{intro}, contradictory conclusions were drawn on the relationship between the DIBs and the UV {\it bump}. We have used the compilation of \cite{Xiang17} and the 3D extinction maps in an attempt to re-examine this point. The compilation has been initiated by \cite{Xiang11}, who concluded on the absence of a correlation between reddening-normalized DIBs and the reddening-normalized {\it bump} strength, based on poor Pearson correlation coefficients, Pr. The strength of the bump was defined as the area above the linear profile in the classical description of \cite{Fitzpatrickmassa90}, namely $\pi c_{3}\gamma^{-1}$ where $c_{3}$ and $\gamma$ are strength and width coefficients of the Drude profile fitted to the bump. The authors used the E($\lambda$-V)/E(B-V) formulation of the extinction law, in the same way as \cite{Desert95}. Later on, and based on an augmented compilation of DIBs and extinction law parameters, \cite{Xiang17} used the A($\lambda$)/$A$($V$) form of the extinction law and defined a new decomposition of the law. Based on this new definition, the authors confirmed the absence of correlation between extinction-normalized DIBs and the newly defined bump area. Here, we re-examined the dependencies on the bump in two parallel ways, using reddening-normalized DIBs and the E($\lambda$-V)/E($B$-$V$) formulation of the extinction law on the one hand (hereafter FORM1) , and extinction-normalized DIBs and the $A$($\lambda$)/$A$($V$) form of the extinction law on the other hand (hereafter FORM2). The bump height is proportional to $c_{3}\gamma^{-2}$ in both cases, but $c_{3}$ is not defined in the same way. 

%We note that, in Fig.~2 from \cite{Xiang11}, showing dependencies of the reddening-normalized DIBs on the bump area, interestingly all linear fits have a positive slope, which statistically should not be the case for a total absence of link. Moreover, the Pearson correlation coefficients, although not very high (ranging from 0.2 to 0.58), suggest a weak to moderate correlation, especially in view of the large number of data points available for the strong DIBs, 
\begin{table*}
\caption{Line-of-sight flags adopted for stars from the \cite{Xiang17} catalogue.}
\label{star_flag}
\begin{tabular}{llllllllll} \hline\hline
star & flag & star & flag & star & flag & star & flag \\ \hline
HD2905 &  0  & HD42087 &  0  & HD143018 &  1  & HD152247 &  0  & HD199216 &  0 \\
HD15558 &  0  & HD46056 &  1  & HD143275 &  1  & HD152248 &  0  & HD199478 &  1 \\ 
HD15570 &  0  & HD46150 &  1  & HD143567 &  1  & HD152249 &  0  & HD199579 &  1 \\ 
HD15629 &  0  & HD46202 &  1  & HD144217 &  1  & HD154445 &  1  & HD200775 &  1 \\ 
HD16691 &  0  & HD46223 &  1  & HD144470 &  1  & HD162978 &  1  & HD203938 &  1 \\ 
HD21291 &  0  & HD47129 &  1  & HD145502 &  1  & HD164492 &  1  & HD204827 &  1 \\ 
HD21483 &  1  & HD47240 &  1  & HD145554 &  1  & HD164794 &  1  & HD206165 &  1 \\ 
HD23060 &  0  & HD48099 &  1  & HD146001 &  1  & HD165052 &  1  & HD206267 &  1 \\ 
HD27778 &  1  & HD48434 &  0  & HD146029 &  1  & HD167971 &  1  & HD207198 &  1 \\ 
HD30614 &  1  & HD53974 &  0  & HD146416 &  1  & HD168076 &  1  & HD209339 &  1 \\ 
HD+31643 &  1  & BD+60497 &  0  & HD147165 &  1  & HD168112 &  1  & HD210121 &  1 \\ 
HD34078 &  0  & BD+60594 &  0  & HD147701 &  1  & HD183143 &  0  & HD216532 &  0 \\ 
HD36879 &  0  & HD99872 &  1  & HD147888 &  1  & HD185418 &  0  & HD216898 &  0 \\
HD37367 &  1  & HD122879 &  1  & HD147889 &  1  & HD190603 &  1  & HD217086 &  0 \\ 
HD37903 &  1  & HD123008 &  0  & HD147933 &  1  & HD192281 &  1  & HD229196 &  1 \\ 
HD38087 &  1  & HD142096 &  1  & HD149757 &  1  & HD193322 &  1  & HD239729 &  1 \\ 
HD38131 &  0  & HD142165 &  1  & HD152233 &  0  & HD193682 &  0  & HD242908 &  0 \\ 
HD40893 &  0  & HD142315 &  1  & HD152236 &  0  & HD197770 &  0  & HD303308 &  0 \\ 
HD41117 &  0  & HD142378 &  1  & HD152246 &  0  & HD198478 &  1  & BD+631964 &  0 \\ 
\hline
\end{tabular}
\end{table*}

As we already discussed, line of sight integrated data can not give rise to tight correlations, due to cloud multiplicity. Our goal was to use the LOS structure as it shows up in the 3D dust map to infer the potential level of degradation of a DIB-bump dependence due to the existence of clouds with different properties. For all targets from the \cite{Xiang17} compilation, we inspected visually the series of clouds along the line of sight using the image of the extinction in a vertical plane containing the Sun and the target, computed from the 3D extinction distribution of \cite{Vergely22}. We, admittedly somewhat arbitrarily, defined a flag F according to the cloud distribution between the Sun and the star. If most of the extinction is produced by a single compact cloud or by a cloud group with limited extension (say, smaller than $\simeq$ 100~pc), then F = 1. If, on the contrary, the extinction is produced by series of clouds distributed over long distances or to two groups separated by a large distance (say, more than $\simeq$ 500~pc), F = 0. This produces two groups of lines of sight of about the same size. The series of targets and their attributed flags is listed in Table \ref{star_flag}. Two examples are shown in Fig.~\ref{visu_flags}, for the target star HD\,192216 ($l$ = 88.92$\degr$, $b$ = 3.02$\degr$, $d$ = 1100~pc) as an example of a flag F = 0, and for HD\,143567 ($l$ = 350.87$\degr$, $b$ = 22.68$\degr$, $d$ = 135~pc) as an example of a flag F = 1. 

\begin{figure}
   \centering
    \includegraphics[width=0.49\textwidth]{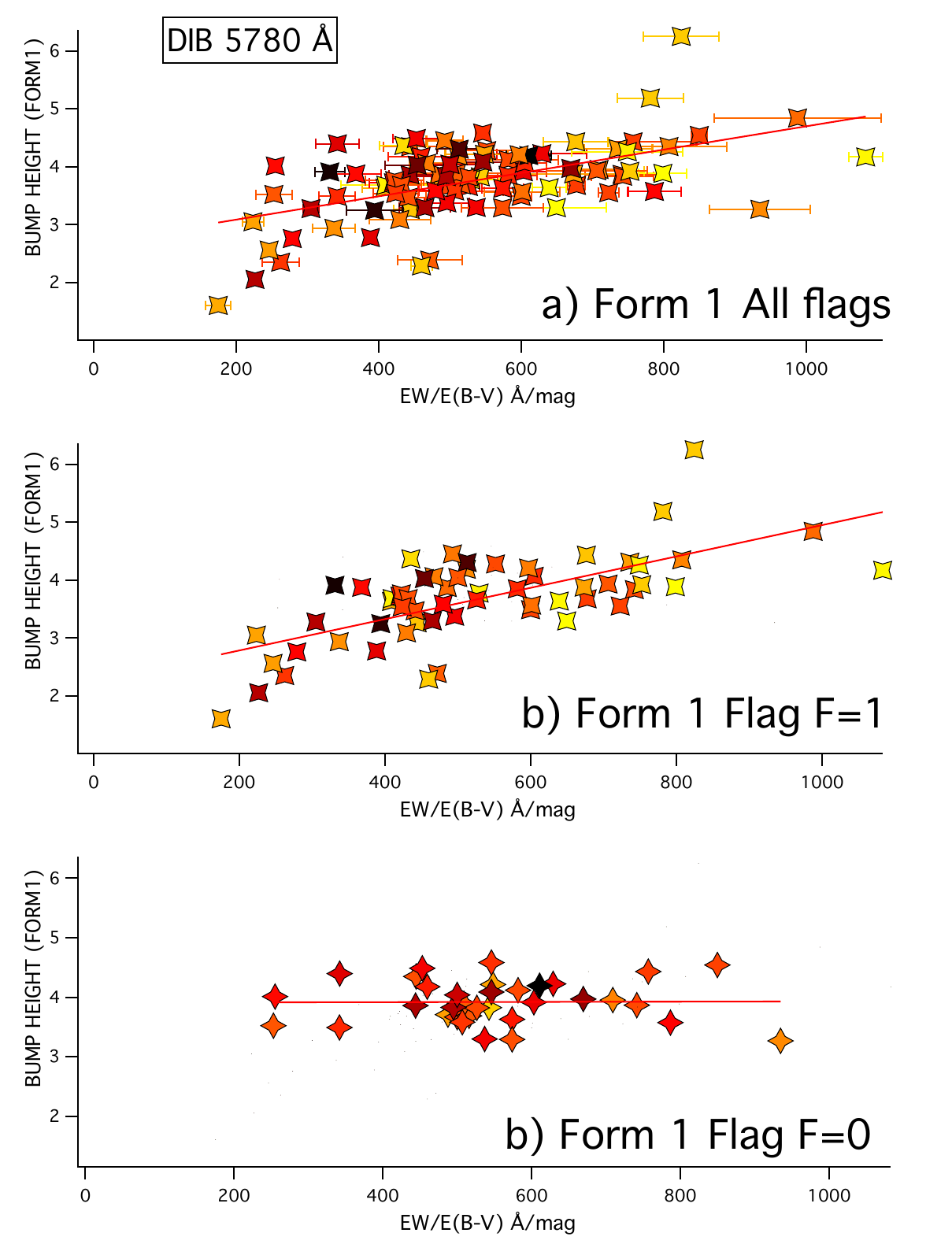}
       \includegraphics[width=0.49\textwidth]{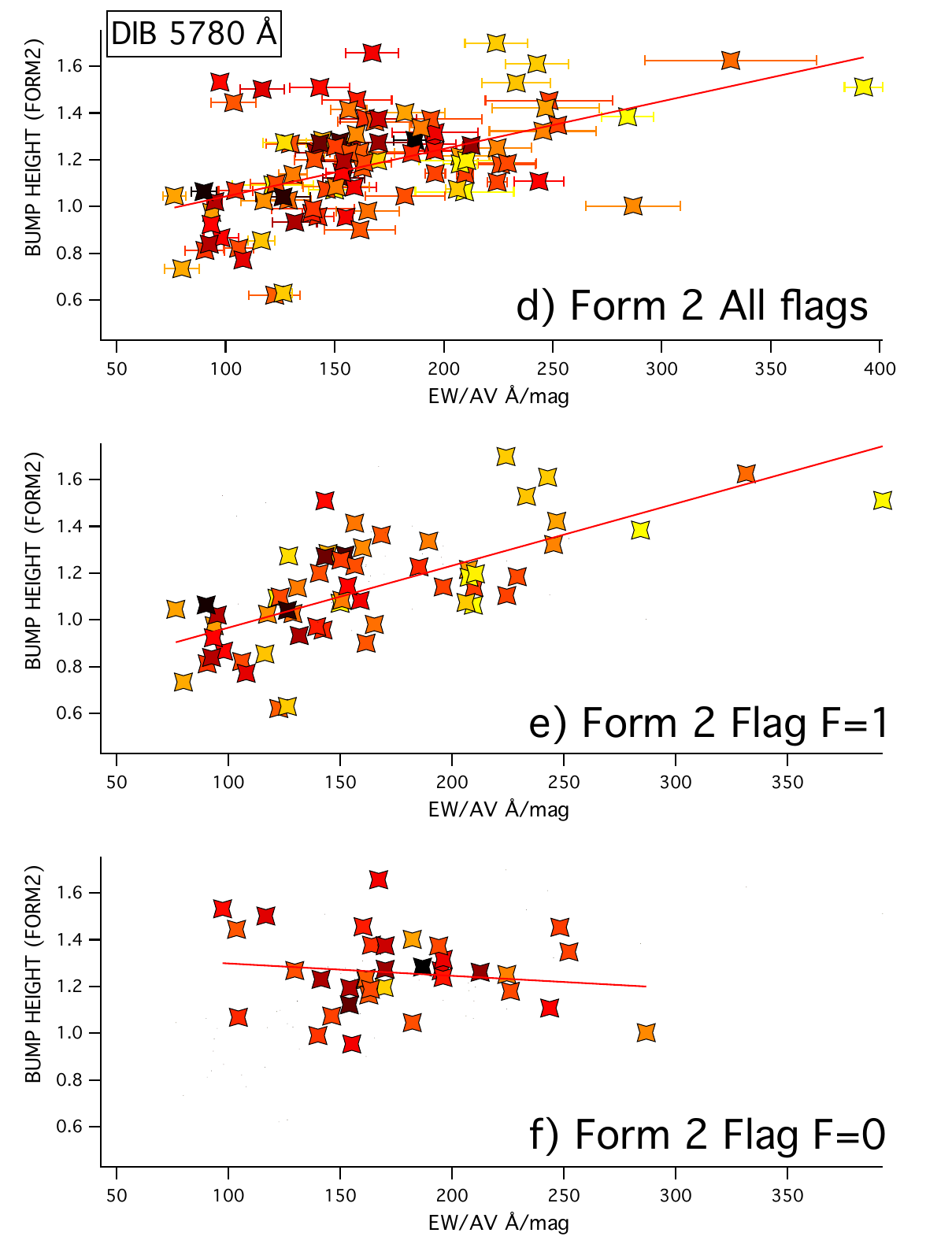}
\caption{Comparisons between the $\sigma$-type 5780 ~\AA\ DIB EW and the UV bump height, under formalism 1 of the extinction law (a,b,c) and formalism 2 (d,e,f). Data are from the compilation of \cite{Xiang17}. a) and d) show the bump height vs the reddening-normalized EW for all targets and the adjusted linear fit. The color code refers to the extinction (from yellow for the lowest extinctions to black for the highest ones). Errors bars on EWs are indicated. b) and e): Same as a) and d), for flag F=1 stars only. c) and f): same as a) and d) for flag F = 0 only. The F=0 lines-of-sight show no dependence, and lower the correlation based on all data. F=1 targets show a clearer positive correlation.}
    \label{dib5780_bump_height_role_LOS}
\end{figure}

\begin{figure}
   \centering
    \includegraphics[width=0.49\textwidth]{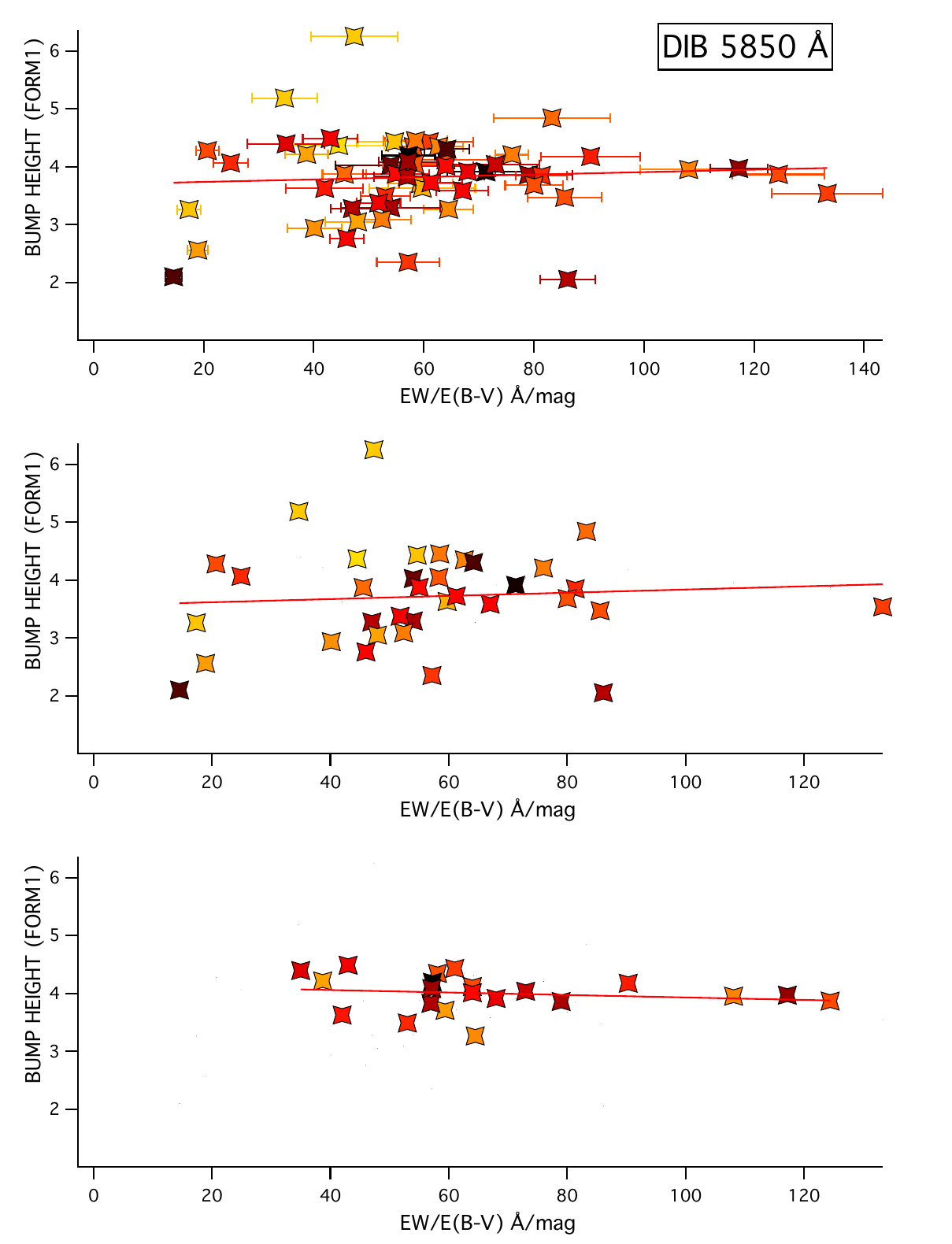}
\caption{Same as Figure \ref{dib5780_bump_height_role_LOS} a),b),c), for the $\zeta$-type 5850 ~\AA\ DIB and formalism 1. Top: all flags. Middle: flag F=1. Bottom: flag F=0. No trend is detected, neither for all stars, nor for F=1 stars only.}
     \label{dib5850_bump_height_role_LOS}
\end{figure}

%\begin{figure}
%   \centering
% %   \includegraphics[width=0.95\textwidth]{DIBNORM_VS_BETA_TWOBETAS_AVINTER}
%    \includegraphics[width=0.99\textwidth]{DIB5780_TWOFORMS_ROLE_LOS_HEIGHT.pdf}
%   \caption{Comparisons between the $\lambda\lambda$5780 DIB EW and the UV bump height. Data are from the compilation of  \cite{Xiang17}. Left: reddening normalized EWs and UV bump height for FORM1 of the extinction law (see text). Right: extinction normalized EWs and UV bump height for FORM2 of the extinction law. Colors refer to the LOS extinction. Marker sizes distinguish Flag 1 (bigger size), and Flag 0 (small size). Superimposed linear fits correspond to Flag 1 only.}
%    \label{dib5780_bump_height_role_LOS}
%\end{figure}

\subsection{Bump height}
We first selected the strong 5780~\AA\ DIB. In Fig.~\ref{dib5780_bump_height_role_LOS} are displayed the reddening- (respectively extinction-) normalized EWs as a function of the bump height in the FORM1 (respectively FORM2) representations of the extinction law. Adjusted linear fits are shown for all stars (top), flag F = 1 stars (middle) and flag F=0 stars (bottom). Uncertainties on the DIB equivalent widths from \cite{Xiang17} are shown at top. The displayed fitted line does not take any error bar into account, errors on the reddening (or on the extinction) being too uncertain. Instead, the dispersion around the fitted line is used as an estimate of the mean prevailing uncertainty. For this particular DIB, the Pearson correlation coefficients are 0.54 $\pm$ 0.09, 0.67 $\pm$ 0.10 and 0.01 $\pm$ 0.17 respectively for 94, 59 and 35 stars for the first formalism. The estimated errors on the correlation coefficient are based on Student's {\it t}-distributions (i.e. assuming normal distributions for input data). In case of the second formalism, these coefficients become 0.51 $\pm$ 0.09, 0.69 $\pm$ 0.10, -0.14 $\pm$ 0.17. The slopes and one $\sigma$ uncertainties are 2.0 $\pm$ 0.3 10$^{-3}$, 2.7 $\pm$ 0.4 10$^{-3}$, 0.02 $\pm$ 0.4 10$^{-3}$ mag\,\AA$^{-1}$ respectively  for formalism 1, and 2.0 $\pm$ 0.3 10$^{-3}$, 2.6 $\pm$ 0.4 10$^{-3}$, -0.5 $\pm$ 0.6 10$^{-3}$ mag\,\AA$^{-1}$ respectively for formalism 2. These results demonstrate that, in both cases, the selection of lines of sight closer to the ideal case of a mono-cloud situation has a strong effect. For LOS characterized by cloud multiplicity (F = 0), the UV {\it bump} height is about constant and there is a reduced dispersion around the linear fit, which we interpret as a consequence of the averaging of the cloud properties. This series of data produces the central {\it plateau} in the figure displaying all assembled data. On the contrary, restricting to those LOS closer to the mono-cloud situation, there is a pronounced positive slope and a large variability around the linear fit. This shows the actual existence of a positive correlation between the strength of this particular DIB and the bump height.

%la tendance a la baisse de correlax entre 5780 et 5850 subsiste 
%donc tout faire en parallele et a chque fois montrer le role des bonnes visees xxxxxxxxxx

%\begin{figure}
%   \centering
%     \includegraphics[width=0.65\textwidth,height=12cm]{bump_3dibs.pdf}
%    \caption{Estimates of the 2200 Ang UV 'bump'  height based on three individual DIBs as a function of the measured bump.}
%    \label{bump}
%\end{figure}
%
%\begin{figure}
%   \centering
%     \includegraphics[width=0.85\textwidth,height=10cm]{BUMP_PREDICT.pdf}
%    \caption{Composite model for the 2200 Ang UV 'bump'  height  based on DIBs and reddening, as a function of the actual, measured bump height.}
%    \label{bump_model}
%\end{figure}

While re-examining in the same way the potential relationships with the bump height for other DIBs than the 5780 ~\AA\ band, we surprisingly found a decreasing correlation strength from the $\sigma$-type DIBs such like the 5780~\AA\ DIB and the 6284~\AA\ DIB to $\zeta$-type DIBs such like the 5797~\AA\  DIB or the 5850~\AA\ DIB. The former ($\sigma$) bands are well known for being significantly reduced in dense, molecular clouds, while the latter do not. The $\sigma$ to $\zeta$ ratio 5780/5797 is often used as an indicator of this difference, and shown to increase with the UV radiation field prevailing in diffuse clouds \citep[e.g.][] {Krelowski87,Cami97}.

%xxx here show same study and figures as above for but 5850 
%xxvoir aussi 6284 je transpose les erreurs relatives en ew en erreurs relative en height puis fitte je trouve sur 41 points un sigma sur H de 0.51 
%pas si mal le mieux est 0.45 avec la 5850 33 pts
\begin{table*}[h!]
\caption{Reddening-normalized DIB EW and {\it bump} height (formalism 1). The number N of lines of sight, the Pearson correlation coefficient Pr, and the uncertainty on the coefficient err(Pr) (see text) are listed. }
\label{bump_correl_form1}
\begin{tabular}{lllllllllll} \hline\hline
DIB (\AA) & N (ALL)  & Pr & Err(Pr) & N (F = 1) & Pr & Err(Pr) & N (F = 0) & Pr  & Err(Pr)\\ \hline
5780  & 94  & 0.54 & 0.09 & 59 & 0.67 & 0.10 & 35  & 0.01 & 0.17 \\ 
6284 & 64   & 0.52 & 0.11 & 41 & 0.57 & 0.13 & 23 & 0.26 & 0.21\\ 
6196 & 76  & 0.50 & 0.10  & 51   & 0.56 & 0.12 & 25  &  0.26 & 0.20\\ 
6614 & 58   & 0.56 & 0.11 & 44 & 0.63 & 0.12 & 14 & 0.16 & 0.28 \\ 
6269 & 59   & 0.32 & 0.12 & 36 & 0.37 & 0.16 & 23 & 0.17 & 0.22\\ 
5797 & 90   & 0.16 & 0.11 & 57 & 0.17 & 0.13 & 33 & -0.02 & 0.18 \\ 
6379 & 43   & 0.25 & 0.15 & 34 & 0.27 & 0.17 & 9 & 0.06 & 0.38\\ 
5850 & 56   & 0.07 & 0.14 & 35 & 0.08 & 0.17 & 21 & -0.17 & 0.22\\ 
\hline
\end{tabular}
\end{table*}

\begin{table*}[h!]
\caption{Extinction-normalized DIB EW and {\it bump} height (formalism 2). The number N of lines of sight, the Pearson correlation coefficient Pr, and the uncertainty on the coefficient err(Pr) (see text) are listed. }
\label{bump_correl_form2}
\begin{tabular}{lllllllllll} \hline\hline
DIB (\AA) & N (ALL)  & Pr & Err(Pr) & N (F = 1) & Pr & Err(Pr) & N (F = 0) & Pr  & Err(Pr)\\ \hline
5780  & 94  & 0.51 & 0.09 & 59 & 0.69 & 0.10 & 35  & -0.14 & 0.17 \\ 
6284 & 64   & 0.51 & 0.11 & 41 & 0.56 & 0.13 & 23 & 0.27 & 0.21\\ 
6196 & 76  & 0.58 & 0.09  & 51   & 0.66 & 0.11 & 25  &  0.20 & 0.20\\ 
6614 & 58   & 0.66 & 0.10 & 44 & 0.73 & 0.10 & 14 & 0.10 & 0.29 \\ 
6269 & 59   & 0.42 & 0.12 & 36 & 0.44 & 0.15 & 23 & 0.26 & 0.21\\ 
5797 & 90   & 0.34 & 0.10 & 57 & 0.41 & 0.12 & 33 & -0.13 & 0.18 \\ 
6379 & 43   & 0.41 & 0.14 & 34 & 0.42 & 0.16 & 9 & 0.29 & 0.36\\ 
5850 & 56   & 0.29 & 0.13 & 35 & 0.35 & 0.16 & 21 & -0.13 & 0.23\\ 
\hline
\end{tabular}
\end{table*}

We show in Fig.~\ref{dib5850_bump_height_role_LOS}, as an example, the results for the $\zeta$-type 5850 ~\AA\ DIB. The relationship between the reddening-normalized DIB EW and the bump height is dramatically different from the one of the 5780 ~\AA\ DIB, since no trend is detectable, whatever the LOS selection, in formalism 1, and a very weak trend is detected for formalism2 (not shown here). 
Tables~\ref{bump_correl_form1} and~\ref{bump_correl_form2} list the correlation coefficients of the adjusted linear fits for the dependence of the bump height on reddening-normalized and extinction-normalized DIB EW. We limited the choice of DIBs to those with more than 12 lines-of-sight. Coefficients are also listed for flag 1 only and flag 0 only. The estimated standard error on the correlation coefficient based on Student's {\it t}-distribution (i.e. assuming normal distributions) is also listed. The DIBs are listed by decreasing Pearson correlation coefficient computed for all flags 1 and 0 LOS and formalism1, i.e. from 0.54 down to 0.07. We also see that the improvement (respectively degradation or total disappearance) of the correlation when considering only flag 1 (respectively flag 0) we found for the 5780 ~\AA\ DIB is also found for all other DIBs, which reinforces the case for actual correlations. We note that all correlations are stronger for formalism 2. However, for both formalisms, the sequence of DIBs for decreasing trends shares similarities with the hierarchy of DIBs associated with the $\sigma$ to $\zeta$ types, or to the PC2 components of \cite{Ensor17}, found by the authors to represent the UV-field and taken from their Fig.~7. 
%(DIB wavelength; Pearson Pr) :
%($\lambda$6284; 0.52 ) ($\lambda$6270; 0.32) ($\lambda$5780; 0.57) ($\lambda$6196; 0.50) ($\lambda$6614; 0.49) ($\lambda$6379;0.27) ($\lambda$5797; 0.25) ($\lambda$5850; 0.07).

\begin{figure*}
   \centering
     \includegraphics[width=0.8\textwidth]{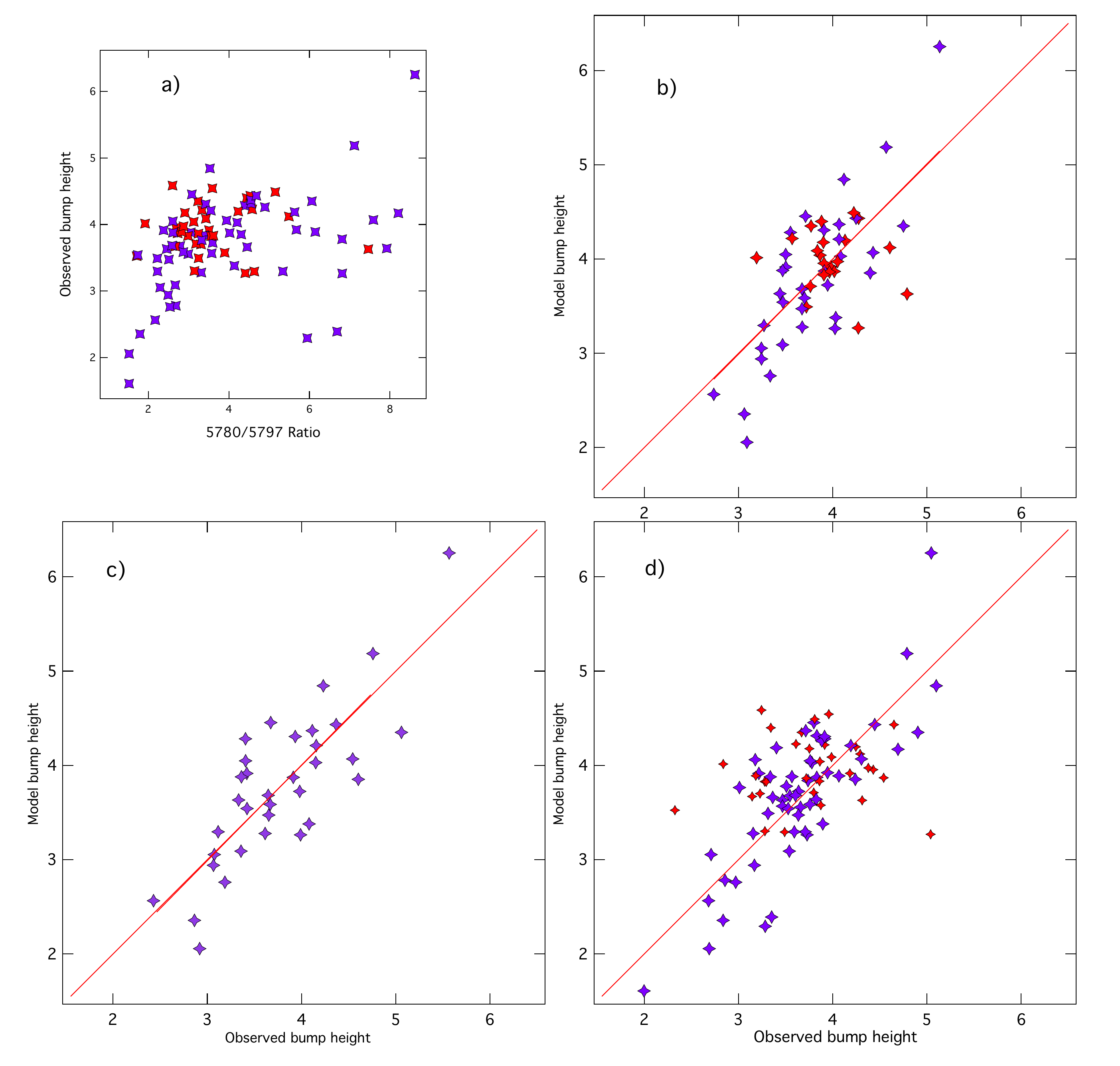}
    \caption{Illustration of a composite model of the bump height based on DIBs and the {\it UV ratio}. a): Bump height and UV ratio 5780/5797.  b): Composite power-law model (Eq.~\ref{compomodel}) based on the 5850 ~\AA\ DIB and the UV ratio, for all stars with DIB 5850~\AA\ data (55 targets). c) : same as Middle, for a model based on Flag 1 stars only. d): Composite power-law model (Eq.~\ref{compomodel}) based on DIB 5780, DIB 5850 and the UV ratio, adjusted on Flag 1 stars (89 targets). Red color is for Flag 0 stars, violet for F = 1. On b,c,d, y=x curves are superimposed.}
    \label{bumpheight_uvratio_5850}
\end{figure*}

\subsection{Estimates of bump height based on DIBs}

The above results suggest that the response of a particular DIB to environmental conditions, here, the strength of the UV radiation field, modifies the link with the bump height. We introduced a {\it UV factor} defined by the classical EW(DIB 5780)/EW(DIB 5797) ratio as an additional factor of the relationship between the bump and the reddening-normalized DIB. We first show in Figure \ref{bumpheight_uvratio_5850} (a) the relationship between the bump height and this {\it UV factor}. For Flag 0 stars there is no trend. For Flag 1 stars there may be some link, visible for extreme low and high values of the height, however, obviously a link far from a linear relationship. We then fitted the UV bump height, i.e. the $c_{3}\gamma^{-2}$ coefficient taken from the \cite{Xiang17} Table~1, to the product of two power laws, one on the reddening-normalized DIB and the second on the {\it UV factor} , i.e. the $\lambda$5780/$\lambda$5797 ratio. 

\begin{equation}
\mathrm{Height(model)}=c_{3}/\gamma^{2} = K\, (\mathrm{DIB_{norm}})^{A1} \cdot (EW5780/EW5797)^{A2} 
\label{compomodel}
\end{equation}

Figure~\ref{bumpheight_uvratio_5850} shows the fitted {\it model} bump height for the case of the 5850~\AA\ DIB, based on the use of all 5850~\AA\ data with Flags 0 and 1 (part b), and based on Flag 1 stars only (part c). The $K$, $A1$ and $A2$ coefficients are 0.35, 0.16, 1.3 and 0.45, 0.23, 0.88 respectively. 
Although this is certainly a far from perfect representation, we can see that this simplistic model reproduces the high and low values of the bump. 
%In this figure, the model bump heights are displayed as a function of the measured bump height from three different DIBS, 5780, 6614 and 5850. The Pearson correlation coefficients are now 0.60, 0.64 and 0.69 respectively. On the other hand, we see that the amplitude of the modeled bump height is significantly smaller than the actual, measured amplitude. 
% The coefficients c1 and c2 are (0.045 ; 0.188), (0.271; 0.264), (0.650 ; 0.723) respectively. 
% Interestingly, the best modeling of the bump height based on DIBs is for the DIB 5850, particularly sensitive to the UV field and reaching strong EWs in UV-protected regions. However, the modeled bump height range is lower than the measured one. 

We pursued this exercise for other DIBs, namely the 5780, 5797, 6196, 6614~\AA\ DIBs. The results were approximately the same. We also tested combinations of power laws for several DIBs. The advantage is to increase the number of targets for the fitting and the comparison, since EWs are not available for all bands. Finally, what comes out from these various attempts is that $\sigma$ DIBs such like 5780~\AA\ can be used directly, they are a kind of reference, while $\zeta$ DIBs, while corrected for their specific UV ratio ($\sigma/\zeta)$), may also provide some estimates of the {\it bump} height. As an example, Fig.~\ref{bumpheight_uvratio_5850} d) shows the adjusted model based on a combination of the 5780~\AA\ and 5850~\AA\ DIBs. Its formulation is: 
\begin{equation}
\begin{aligned}
\mathrm{Height(model)} = 0.58 \cdot (\mathrm{DIB(5780)}_\mathrm{norm})^{0.27} \cdot (5780/5797)^{0.073} \\ 
\cdot (\mathrm{DIB(5850)}_\mathrm{norm})^{0.026} %\\
%Mod_{T}= (\Sigma Mod_{i} , i=1, 5) / 5 \\
%ModF_{T} =( (Mod_{T}-3.75) \cdot 2.3 ) + 3.75 
\end{aligned}
\end{equation}

%MODIFY xxx The final model ModF$_{T}$ is then a function of the reddening and of the five EWs of the $\lambda$5780, 5797, 5850 , 6196 and 6614 DIBs. Fig. \ref{bump_model} shows the bump height prediction based on DIBs and EBV as a function of the measured bump, for the 89 lines of sight with available measurements of the extinction law and at least one of the 5 DIBs others than 5780 and 5797. Although far from perfect, this modeling shows that some constraints on the bump height can be made on the basis of the DIBs, especially in the case of extreme (low and high) values of the bump (and the model). 

%xx modify in fact it is even better to use only 5850 as a dib and 5780/5797 for the ratio
%show the results maybe the log representation showing the proportions of both factors
This simplistic model reproduces the high and low values of the bump, and may provide a prior coefficient for the bump height. The 1$\sigma$ deviation from the model is 0.5, to compare with height values ranging from 2 to 5. Further extinction law measurements would help refining and correcting this approach purely based on DIBs. 

\subsection{Bump width and bump area}

\begin{figure}
   \centering
   \includegraphics[width=0.4\textwidth]{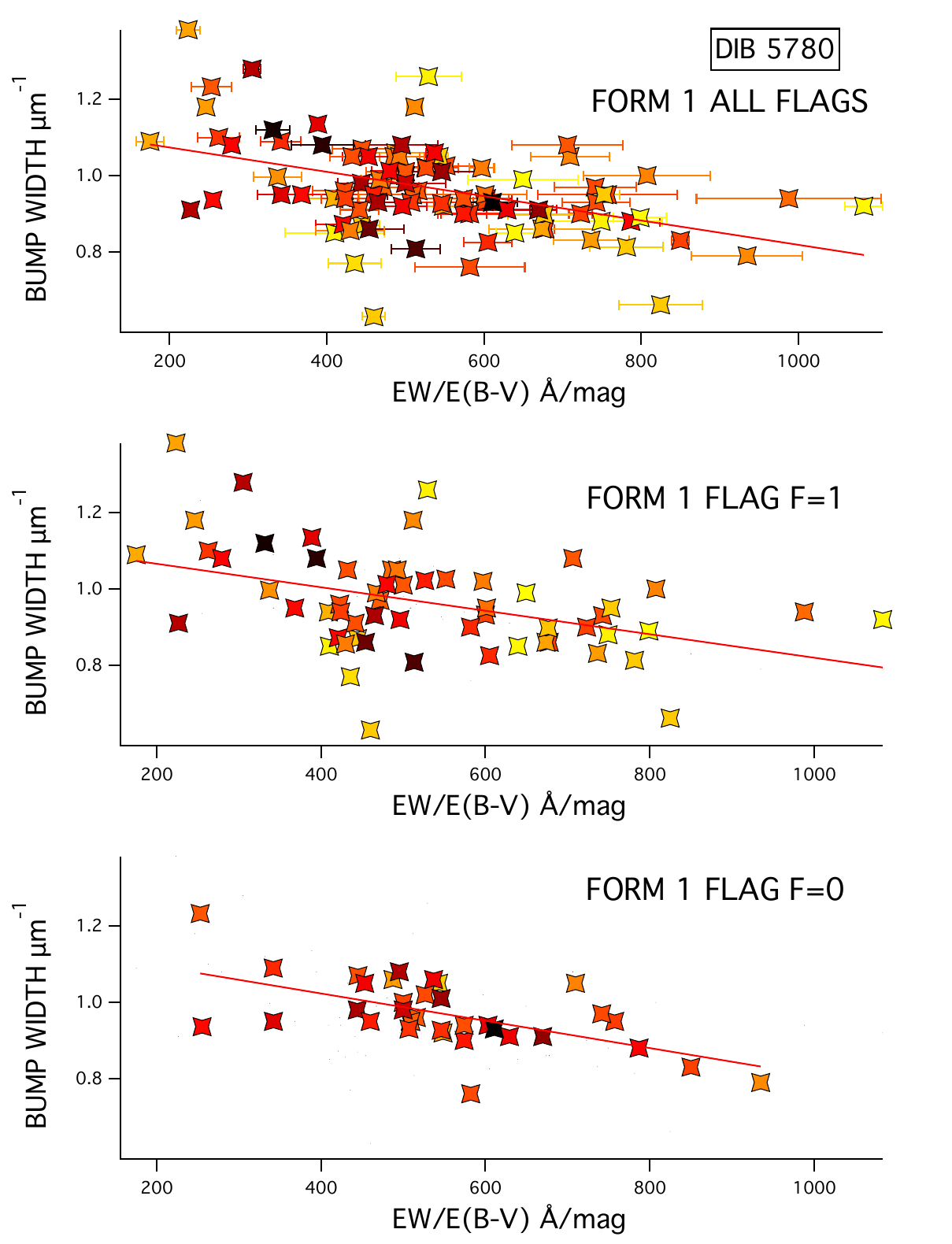}
   \includegraphics[width=0.4\textwidth]{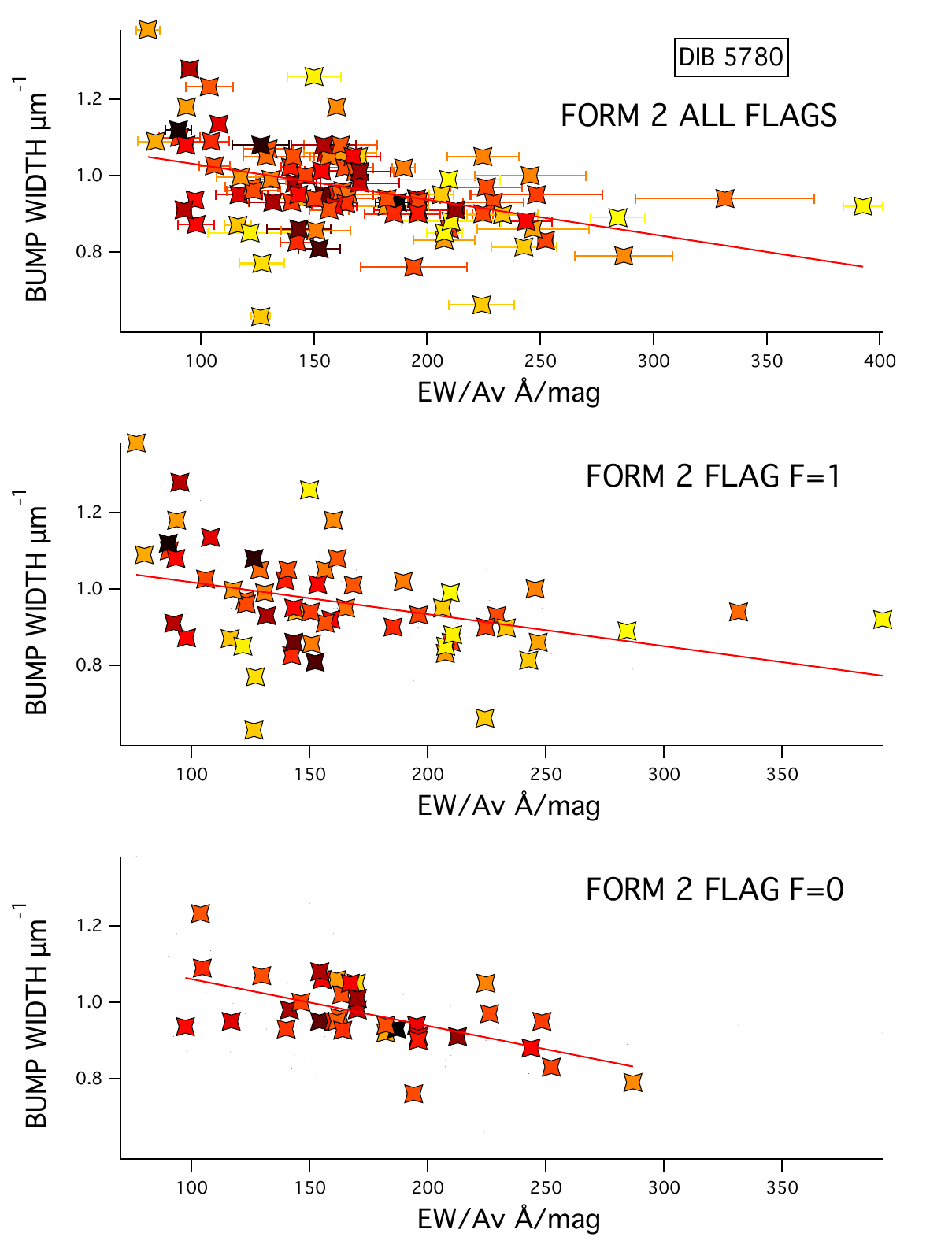}
   \caption{Same as Figure \ref{dib5780_bump_height_role_LOS} for the dependence on the bump width, for formalism 1 and 2. Top: all targets. Middle: Flag 1 targets. Bottom Flag zero targets.}
    \label{dib5780_bump_width_role_LOS}
\end{figure}

We made the same comparisons and linear fits, this time as a function of the bump width and the bump area. The correlations with the bump width are displayed in Fig.~\ref{dib5780_bump_width_role_LOS} for the strong 5780~\AA\ DIB. The Pr factors do vary from -0.46$\pm$0.09 to -0.42 $\pm$ 0.12 and reduced $\chi^2$ from 8.9 to 7.4 from before and after LOS selection for formalism 1. Pr factors are almost constant at -0.42 $\pm$ 0.09 and -0.38 $\pm$ 0.12 and reduced $\chi^2$ constant at 7.4 for formalism 2. The slopes and their 1$\sigma$ uncertainties are -3.2 $\pm$ 0.6 10$^{-4}$, -3.1 $\pm$ 0.9 10$^{-4}$ , -3.6 $\pm$ 0.8 10$^{-4}$ $\mu$m${-1}$ mag A$^{-1}$ respectively for 94, 59, 35 targets with FLAGS 0 and 1, 1 , and 0 for formalism 1. The numbers become -9 $\pm$ 2 10$^{-4}$, -8.4 $\pm$ 3 10$^{-4}$ , -12.2 $\pm$ 3 10$^{-4}$ $\mu$m${-1}$ mag A$^{-1}$ for formalism 2. The uncertainties on the slopes are larger than in the case of the bump height, but all results favor an inverse trend between the reddening normalized DIB and the width. 
Results for correlation factors and uncertainties, for the same DIBs and in the same order as in Table~\ref{bump_correl_form1} for the height are listed in Table~\ref{width_correl}. We listed the results for formalism 1 only. Results for formalism 2 are similar. There is again a net decrease of the correlation (here anti-correlation) strengths from $\sigma$ to $\zeta$ DIBs. However, at variance with the case of the height, the strength of the correlation for the width does not seem to be linked to the LOS type. 

\begin{figure}
   \centering
    \includegraphics[width=0.49\textwidth]{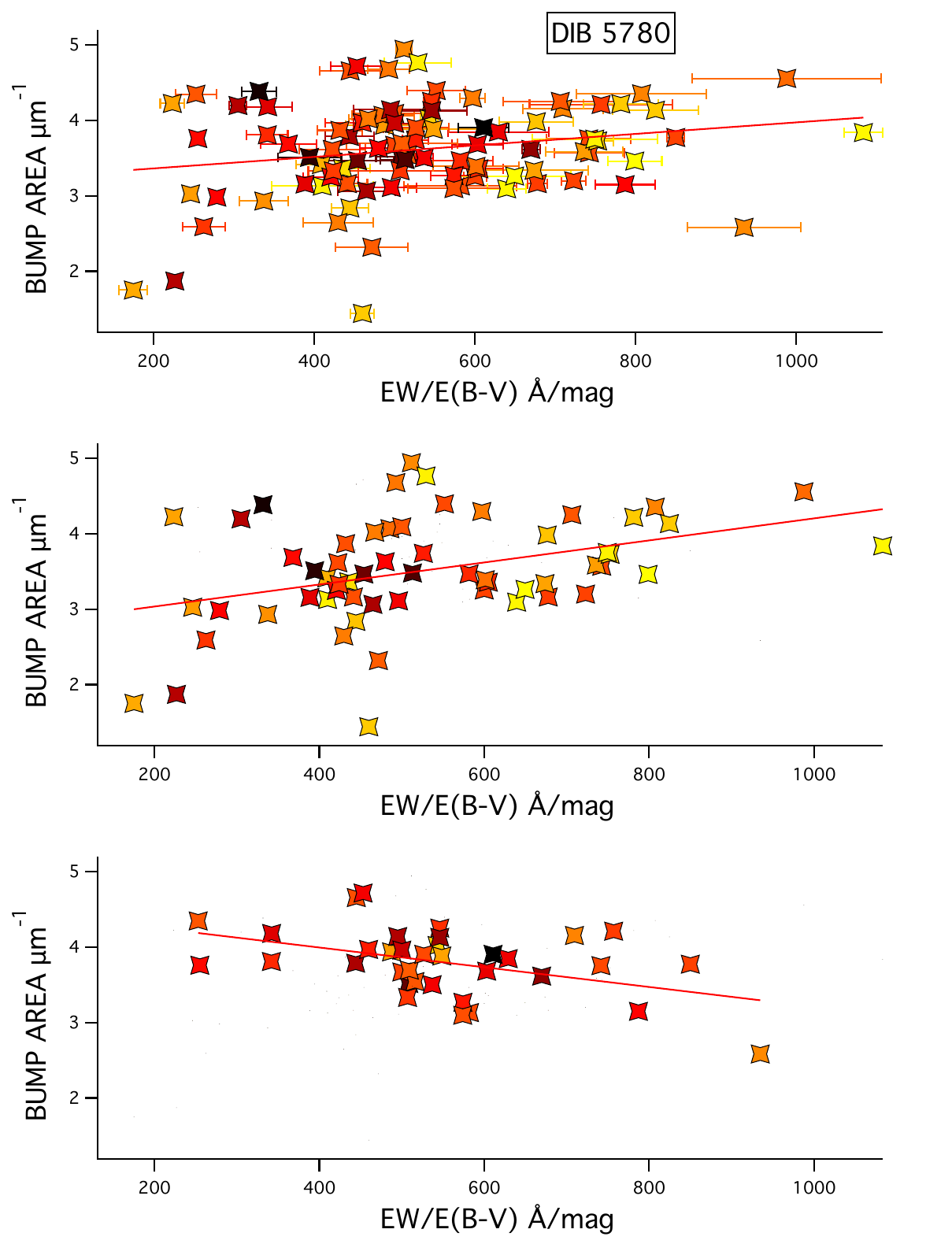}
   \caption{Same as Figure \ref{dib5780_bump_height_role_LOS} for the dependence on the bump area and formalism 1. Top: all flags. Middle: F=1. Bottom: F=0.}
    \label{dib5780_bump_area_role_LOS}
\end{figure}

The results for the bump area are shown in the case of formalism 1 in Fig.~\ref{dib5780_bump_area_role_LOS}. The slopes and their 1$\sigma$ uncertainties are 7.6 $\pm$ 3.7 10$^{-4}$, 14 $\pm$ 5 10$^{-4}$ , -13 $\pm$ 5 10$^{-4}$~mag~$\mu$m$^{-1}$~\AA$^{-1}$ respectively for 94, 59, 35 targets with FLAGS 0 and 1, 1 , and 0 for formalism 1. The Pearson correlation coefficients are 0.21 $\pm$ 0.10, +0.39 $\pm$ 0.12, and -0.44 $\pm$ 0.16 respectively for the 94, 59 and 35 targets. The uncertainties on the slopes of the linear fits are much larger than in the case of the bump height and still significantly larger than for the bump width. Indeed, at variance with the case of the bump height, and the bump width, there is no constancy of the slope sign, since there are two opposite trends for Flag1 and Flag0, and the links are unclear. 
Results for correlation factors and uncertainties, for the same DIBs and in the same order as in Table~\ref{bump_correl_form1} for the height are listed in Table~\ref{area_correl}. 

All together, the results are in agreement with those from \cite{Desert95} for reddening-normalized DIBs, namely a positive correlation with the bump height and a negative correlation with the bump width, provided one uses $\sigma$-type DIBs and mono-cloud-type lines-of-sight. They also agree with the absence of a clear global trend for the bump area found by \cite{Xiang17} for extinction-normalized DIBs, and by \cite{Xiang11} for reddening-normalized DIBs. This is consistent with the fact that the area is close to the product of the width and the height. At variance with the results of \cite{Xiang17}, we get some positive correlations for the area, although weaker than for the height, when we restrict to Flag F=1 and if we use $\sigma$-type DIBs. We note that the area and the width are more difficult to measure than the height, in particular, they depend on the choice of the underlying linear part of the extinction law, and this may explain partly the results. 

\begin{table*}[h!]
\caption{Reddening-normalized DIB EW and {\it bump} width. The number N of lines of sight, the Pearson correlation coefficient Pr, and the uncertainty on the coefficient err(Pr) (see text) are listed. }
\label{width_correl}
\begin{tabular}{lllllllllll}  \hline\hline
DIB (\AA) & N (ALL)  & Pr & Err(Pr) & N (F = 1) & Pr & Err(Pr) & N (F = 0) & Pr  & Err(Pr)\\ \hline
5780  & 94  & -0.46 & 0.09 & 59 & -0.42 & 0.12 & 35  & -0.59 & 0.14 \\ 
6284 & 64   & -0.50 & 0.11 & 41 & -0.53 & 0.14 & 23 & -0.41 & 0.20\\ 
6196 & 76  & -0.23 & 0.11  & 51   & -0.18 & 0.14 & 25  &  -0.42 & 0.19\\ 
6614 & 58   & -0.23 & 0.13 & 44 & -0.20 & 0.15 & 14 & -0.72 & 0.20 \\ 
6269 & 59   & -0.10 & 0.13 & 36 & -0.09 & 0.17 & 23 & -0.22 & 0.21\\ 
5797 & 90   & -0.13 & 0.10 & 57 & -0.14 & 0.13 & 33 & -0.13 & 0.18 \\ 
6379 & 43   & -0.06 & 0.16 & 34 & 0.02 & 0.18 & 9 & -0.54 & 0.32\\ 
5850 & 56   & -0.07 & 0.14 & 35 & -0.08 & 0.17 & 21 & -0.03 & 0.23\\ 
\hline
\end{tabular}
\end{table*}

\begin{table*}[h!]
\caption{Reddening-normalized DIB EW and {\it bump} area. The number N of lines of sight, the Pearson correlation coefficient Pr, and the uncertainty on the coefficient err(Pr) (see text) are listed. }
\label{area_correl}
\begin{tabular}{lllllllllll} 
 \hline
DIB (\AA) & N (ALL)  & Pr & Err(Pr) & N (F = 1) & Pr & Err(Pr) & N (F = 0) & Pr  & Err(Pr)\\ \hline
5780  & 94  & 0.21  & 0.10 & 59 & 0.39 & 0.12 & 35  & -0.44 & 0.16 \\ 
6284 & 64   & 0.23 & 0.12 & 41 & 0.32 & 0.15 & 23 & -0.11 & 0.22 \\ 
6196 & 76  & 0.37 & 0.11  & 51   & 0.50 & 0.12  & 25  &  -0.13 & 0.21 \\ 
6614 & 58   &0.43 & 0.12 & 44 & 0.55 & 0.13 & 14 & -0.50 & 0.25 \\ 
6269 & 59   & 0.30 & 0.13 & 36 & 0.39 & 0.16 & 23 & -0.02 & 0.22\\ 
5797 & 90   & 0.10 & 0.11 & 57 & 0.11 & 0.14 & 33 & -0.12 & 0.18 \\ 
6379 & 43   & 0.25 & 0.15 & 34 & 0.35 & 0.16 & 9 & -0.38 & 0.35\\ 
5850 & 56   & 0.08 & 0.14 & 35 & 0.09 & 0.17 & 21 & -0.13 & 0.23 \\ 
\hline
\end{tabular}
\end{table*}

%\begin{figure}
%   \centering
% %   \includegraphics[width=0.95\textwidth]{DIBNORM_VS_BETA_TWOBETAS_AVINTER}
%    \includegraphics[width=0.99\textwidth]{DIB5780_TWOFORMS_ROLE_LOS_AREA.pdf}
%   \caption{Same as Figure \ref{dib5780_bump_height_role_LOS} for the dependence on the bump area.}
%    \label{dib5780_bump_area_role_LOS}
%\end{figure}
%
%\begin{figure}
%   \centering
% %   \includegraphics[width=0.95\textwidth]{DIBNORM_VS_BETA_TWOBETAS_AVINTER}
%    \includegraphics[width=0.99\textwidth]{DIB5780_TWOFORMS_ROLE_LOS_WIDTH.pdf}
%   \caption{Same as Figure \ref{dib5780_bump_height_role_LOS} for the dependence on the bump width.}
%    \label{dib5780_bump_width_role_LOS}
%\end{figure}

\section{Summary and discussion}\label{Conclusion}

The existence of links between DIBs and dust properties has been the subject of numerous works, mainly based on spectra of hot, strongly reddened stars. The results have been limited, and sometimes apparently contradictory. Several new kinds of studies  are now permitted by the availability of 3D maps and massive datasets from {\it Gaia} for the 862 nm DIB, a diffuse band likely representative of most other DIBs : (i) the use of local diagnostics instead of line-of-sight integrated ones, (ii) the use of statistical information on all lines of sight avoiding dense molecular clouds, i.e. low extinction LOS and dust-poor areas, (iii) the cross-match of the information contained in 3D maps with independent information from different surveys, by means of integration in the 3D maps along any line-of-sight. Moreover, for any DIB, the existence of 3D dust extinction density maps permits (iv) the use of the line-of-sight structure to weigh correlative studies. We used these four new possibilities, and, over the course of searches for correlations, we obtained several new results. 

In Sect.~\ref{sect1} we used (i) and (ii) and examined the relative increase of the 862~nm DIB density with respect to the extinction density occurring in low extinction clouds, a trend detected in the 3D maps \citep{Cox24}. We extended the study to line-of-sight integrated data of the two 862~nm DIB catalogs. We fitted the ratio of local (resp. LOS-integrated) DIB and extinction to a power law of the local (resp. integrated extinction) in the low-extinction regime ($A_V$ between 0.12 and 1.5 mag). The fitted power-law coefficients, comprised between 0.57 and 0.74 for LOS data from the two catalogues and for local (3D map) data, were compared with those found by \cite{Lan2015} for a series of 20 DIBs and for the same range of extinctions, based on SDSS stacked spectra for high latitude directions. We found them in agreement with the coefficients of several other DIBs.
This implies that the 862~nm DIB behaves similarly to several other DIBs in the low extinction regime, with a power-law coefficient at the lower end of the \cite{Lan2015} series, i.e. presenting one of the highest relative increases with respect to extinction in dust-poor clouds (lowest values of $\gamma$). This study based on local values and massive data validates the behavior found with SDSS. The variability of the power-law $\gamma$ coefficients found by \cite{Lan2015}, up to now poorly discussed, contains information on the presence of the carriers far from the dense clouds. We note that the hierarchy of DIBs associated with these coefficients is significantly different from the one associated with the component 2 of the Principal Component Analysis of \cite{Ensor17}, assumed to represent the influence of the radiation field. It may be the manifestation of one of the unidentified components in their PCA analysis. Interestingly, the ranking of the 862 nm DIB, based on our results, among the 20 DIBs from the \cite{Lan2015} study, includes a comparison with the broad 443 nm DIB. This comparison implies that the 443 nm DIB density increases more moderately in dust-poor clouds compared to the 862 nm DIB. This is consistent with the recent results of \cite{Zhao23}, who show a smaller extension of the 443 nm DIB carriers at large distance from the Plane in comparison with the 862 nm DIB carriers. This opens the possibility to use the \cite{Lan2015} coefficients to rank the scale heights of the various DIB carriers. 

In Sect.~\ref{sect2} we used (iii) and investigated the link between the 862~nm DIB strength normalized to the extinction on the one hand, and the proxy $R'_V$ for the total-to-selective extinction ratio $R_V = A_V$/E($B$-$V$) derived by \cite{Schlafly16} for a large number of SDSS/APOGEE target stars. One of the results from this work is the existence of large-scale variations of $R_V$ over the sky. For all the targets from the \cite{Schlafly16} catalogue, we integrated from the Sun to the target within the 3D map of 862~nm DIB carrier density on the one hand, and within the corresponding 3D map of extinction density on the other hand (maps from \cite{Cox24}). The results were combined to obtain estimates of the extinction-normalized DIB$_\mathrm{norm}^{862}$. Based on this dataset, and despite a large scatter due to the poor resolution of the 3D DIB map, we detected a positive correlation between DIB$_\mathrm{norm}^{862}$ and $R'_V$ for the low to moderate reddening regime $A_V$ $\lessapprox$ 2.5~mag. Combining with measurements of \cite{Ramireztannus18} and their estimates of $R_V$ for stars in the M17 cloud with $A_V$ $\geq$ $\simeq$ 4~mag, it appears that at higher extinctions the trend is suppressed and reversed. This type of bi-modal regime may explain the absence of previous clear results on the DIB-$R_V$ relationship when data from the two regimes are mixed together. 

In Sect.~\ref{sect3} we also used (iii) and investigated the link between DIB$_\mathrm{norm}^{862}$ and the dust emission spectral index $\beta$. To do so, we used Planck $\beta$ sky maps and selected targets from the two 862 nm catalogs located outside the dust layer, i.e. with negligible interstellar dust beyond the star. This is necessary, since the spectral index $\beta$ refers to the totality of the dust in the chosen direction. Such targets are preferentially found at high latitude. We integrated in the best resolution 3D extinction maps to estimate the extinction of the targets. Despite a large scatter, we found a negative correlation between DIB$_\mathrm{norm}^{862}$ and $\beta$ for $A_V$ $\leq \simeq$ 2.5~mag, and a disappearance of the trend above this threshold, i.e. for high extinctions. There may be hint for reversal, however, the number of appropriate targets with high extinctions and located beyond the dust layer becomes very small. The trend at low-moderate $A_V$ is more apparent when using recent estimates of $\beta$ by \cite{Casandjian22}. 

The two trends, for $R'_V$ and $\beta$, in the low to moderate extinction regime, are found fully consistent with the anti-correlation between $\beta$ and $R'_V$ found by \cite{Schlafly16}, which reinforces the likeliness of these two independent results. The opposite trends we have derived are particularly interesting in view of the models and results of \cite{Zelko20}, namely that the $\beta$ and $R_V$ anti-correlation is induced by variations of the relative fractions of carbonaceous and silicate dust grains, with a higher $R_V$ (and lower $\beta$) being associated with a higher fraction of carbonaceous grains. In this case, because it is now generally admitted that DIB carriers are carbonaceous macro-molecules, their density relatively to the total dust content, measured by DIB$_\mathrm{norm}^{862}$, is {\it naturally} expected to correlate with $R_V$ and anti-correlate with $\beta$, which is what we derive. The trends are no longer valid for lines-of-sight crossing opaque clouds, where accretion onto grains and grain coagulation dominate, and the so-called {\it skin effect} for the DIB carriers starts to play a role. About the carbon to silicon abundance, we note that \cite{Cox24} found evidences from the 3D maps that higher DIB$_\mathrm{norm}^{862}$ values are spatially associated with regions where C-rich ejecta from AGBs increase relatively to the O-rich fluxes. From the point of view of extinction and emission laws, this shows that useful predictions on the dust composition may be obtained based on DIB measurements. 

In Sect.~\ref{sect4} we used (iv) and re-examined the link between the 220~nm UV bump and the DIBs, using data from the intensive compilation of \cite{Xiang17}. The results, up to now, have been contradictory about the existence of a link. One of the reasons for the difficulty of such a study is the cloud multiplicity and the resulting mixing of the cloud (and DIB carrier) properties along the line of sight of the target star, especially in the case of distant, highly reddened hot stars generally used for UV extinction law measurements and DIB extraction. Here we added to the previous searches an additional criterion based on the 3D dust maps and reflecting the cloud {\it singleness} (item iv in the above list). If the LOS is characterized by a unique dense cloud or group of clouds at short distance from each other, the flag is set to 1. If, on the contrary, there are two or more groups of clouds contributing to the extinction (and to the DIB EW) separated by distances on the order of several hundred parsecs or more, one may expect different properties of the IS matter and the flag is set to 0. This criterion is defined visually, and, obviously, is to be used with limited ambition. However, our goal was to detect (or not) an improvement of the link between DIBs and the bump when selecting Flag 1 LOS by comparison with the full dataset or with Flag 0 LOS. This approach allowed us to demonstrate the existence of a positive relationship between the reddening- (or extinction-) normalized DIB EW and the height of the bump for those DIBs that are of $\sigma$ type, known to be favored in UV-irradiated regions and decrease in cloud cores (the UV field here is estimated by the 5780/5797 ratio). It confirmed, based on the larger number of lines of sight from the \cite{Xiang17} compilation, the previous conclusions drawn by \cite{Desert95}. The relationship disappears for DIBs such like $\lambda$5797 or $\lambda5850$, i.e. of $\zeta$ type. This may explain the negative conclusion of \cite{Xiang11} and \cite{Xiang17} about a global DIB-bump relationship associating all DIBs. We tested a simplistic predictive model of the bump height based on a combination of reddening-normalized DIBs and the 5780/5797 ratio. It provides some estimates of the bump height, especially in the case of particularly low, or particularly high bumps. On the other hand, we found a negative relationship between the reddening-normalized 5780 ~\AA\ DIB EW and the bump width, similarly to \cite{Desert95}, although this trend is characterized by a larger scatter and a higher uncertainty on the slope compared to the case of the bump height. Finally, no general trend is emerging from the comparisons between DIBs and the bump area, in agreement with the results of \cite{Xiang17}, although a weak positive trend is found for $\sigma$-type DIBs.

These results on the UV bump are surprising, in particular the fact that direct positive correlations between the bump height and DIBs, and direct anti-correlations between bump widths and DIBs, are found only for DIBs that are the most prone to the {\it skin effect} (or $\sigma$ DIBs), the decrease of the carriers relatively to the extinction in the dense cloud cores. Correlations may also be found in the case of $\zeta$ DIBs, however, only after application of a corrective factor whose effect is to increase their strengths in LOS characterized by a high 5780 ~\AA\ to 5797 ~\AA\ DIB ratio (or $\sigma$ to $\zeta$ ratio).  There may be here evidence for the building of the bump outside of the densest cores, and this point deserves further observations and modeling. In this respect, it is interesting to recall the result obtained by \citet{FitzpatrickMassa86} on the bump width. The authors inferred that the 220 nm bump is narrower in the diffuse ISM, and broader in dense, quiescent clouds. Given that for $\sigma$-type DIBs such like the 5780~\AA\ DIB, DIB$_\mathrm{norm}^{862}$ is found to decrease with the bump width (Fig.~\ref{dib5780_bump_width_role_LOS}), the \cite{FitzpatrickMassa86} trend predicts, in turn, that DIB$_\mathrm{norm}^{862}$ should increase from dense clouds to regions of lower density. This is what is statistically found based on 3D maps, and mono-cloud type LOS, as discussed in Sect.~\ref{sect1}, and adds a consistent link between the trends found between DIB$_\mathrm{norm}^{862}$ and the bump on the one hand, and DIB$_\mathrm{norm}^{862}$ and the ISM type on the other hand.  

It is beyond the scope of this observational work to discuss implications in terms of DIB carriers and species responsible for the UV bump. Our unique remark is that, despite the limited strength of all correlations, as expected from the integration along lines of sight, the results on bump height and width maybe hints of a partial contribution to the bump of some specific molecular species and their specific lines, in addition to the global contribution of a large variety of species, e.g. PAHs. In this respect, it is interesting to note that recent works point to high densities of two-ring poly-cyclic aromatic hydrocarbons, their formation around AGBs and their potential contribution  to the bump. \cite{Stockett23} found a mechanism explaining why cyanonaphthalene (C10H7CN) can be  orders of magnitude higher than predicted by models, following the discovery of quantities of this species and of indene in the Taurus Molecular Cloud with quantities orders of magnitude higher than expected \citep{McGuire21,Burkhardt21}. According to \cite{Stockett23}, their results challenge the widely accepted picture of rapid destruction of small PAHs in space. As a result of their advanced models, \cite{Krasnoukhov22} found that under the conditions prevailing in the circumstellar envelopes of the asymptotic giant branch stars (AGBs), the formation of two-ring poly-cyclic aromatic hydrocarbons is favored. On the other hand, based on state-of-the-art laboratory experimental setup, \cite{Frereux23} found that substituted naphthalene units as substructures of hydrogenated amorphous carbons are favored in combustion processes and produce spectral signatures around 217.5~nm. Further measurements of the bump central wavelength and its potential association with the LOS type may bring support to the existence of such contributions. 
%xxx discuss this from paper by Wang Yang Li MNRAS OCT 2023 they say In general, broad bumps are often seen in dense, quiescent environments, while narrower bumps are mainly seen in diffuse regions and regions of recent early-type star formation (Fitzpatrick \& Massa 1986, hereafter FM86; Decleir et al. 2022; Massa, Gordon \& Fitzpatrick 2022; Gordon et al. 2023).
 
%Are there any consistent links between the various correlations described above? We already discussed the links between DIB$_\mathrm{norm}^{862}$ and $R_V$ in Sect.~\ref{sect2} and between DIB$_\mathrm{norm}^{862}$ and $\beta$ in Sect.~\ref{sect3}, consistent with the $R_V$ and $\beta$ inverse correlation, and likely associated with the carbonaceous fraction. There is another consistency, now between the DIB density (Sect.~\ref{sect1}) and the bump (Sect.~\ref{sect4}). \citet{FitzpatrickMassa86} inferred that the 220 nm bump is narrower in the diffuse ISM, and broader in dense, quiescent clouds. Given that for $\sigma$-type DIBs such like the 578~nm DIB, DIB$_\mathrm{norm}^{862}$ is found to increase with the bump height, and decrease with the bump width, the \cite{FitzpatrickMassa86} trend predicts that DIB$_\mathrm{norm}^{862}$ should increase from very dense clouds to regions of low density. This is what is statistically found based on 3D maps, and mono-cloud type LOS, as discussed in Sect.~\ref{sect1}. 

All together, these results based on 3D maps and surveys reveal complex, but real relationships between the DIB carrier densities and some dust extinction and emission parameters. They suggest that future precise, massive measurements of various DIBs and of extinction, as well as massive measurements of the UV part of the extinction law, may provide constraints on the interplay between the carbonaceous macro-molecules at the origin of DIBs and the size distribution of grains, on the spatial variability of the dust composition, and, in turn, on the variability of extinction and emission law parameters. DIBs are easy to measure and their EWs are calibration-free measurements, these are strong advantages. They are no longer restricted to early-type stars but extended to the far more numerous cool targets \citep[e.g][]{Kos13, Puspitarini15, Zasowski15}, thanks to the huge progresses made over the last years on stellar models and on their adjustments to data from recent surveys and {\it Gaia}. Being distance-limited data, the use of distributed targets provides local measurements, allowing, in turn, to understand where and how are built the extinction and emission, and to make predictions.

\begin{acknowledgements}
This project has received funding from the European Union’s Horizon 2020 research and innovation programme under grant agreement No 101004214 (EXPLORE project – https://explore-platform.eu).\\

This work has made use of data from the European Space Agency (ESA) mission
{\it Gaia} (\url{https://www.cosmos.esa.int/gaia}), processed by the {\it Gaia}
Data Processing and Analysis Consortium (DPAC,
\url{https://www.cosmos.esa.int/web/gaia/dpac/consortium}). Funding for the DPAC
has been provided by national institutions, in particular the institutions
participating in the {\it Gaia} Multilateral Agreement.\\

Funding for the SDSS and SDSS-II has been provided by the
Alfred P. Sloan Foundation, the Participating Institutions, the
National Science Foundation, the U.S. Department of Energy,
the National Aeronautics and Space Administration, the
Japanese Monbukagakusho, the Max Planck Society, and the
Higher Education Funding Council for England. The SDSS
Web Site is http://www.sdss.org/.\\

This research has made use of the VizieR catalog access tool, CDS, Strasbourg, France (DOI: 10.26093/cds/vizier). The original description of the VizieR service is published in 2000, A\&AS 143, 23.

\end{acknowledgements}

\bibliographystyle{aa}
\bibliography{dibemissabsorb}

\end{document}